\documentclass[twocolumn,notitlepage,prl,superscriptaddress,longbibliography]{revtex4-2}
\usepackage{amsfonts}
\usepackage{textcomp}
\usepackage{times}
\usepackage{graphicx}
\usepackage{float}
\usepackage{latexsym,amsmath,amssymb,bm,euscript}
\usepackage{color}
\usepackage{subfigure}
\usepackage{epstopdf}
\usepackage[colorlinks=true,linkcolor=blue,citecolor=blue]{hyperref}
\usepackage{hyperref}
\usepackage{soul}
\usepackage[normalem]{ulem}
\usepackage{mathrsfs}
\usepackage{amsmath}
\usepackage{xspace}
\usepackage{natbib}
\usepackage{ulem}
\usepackage{etoolbox}
\usepackage{tikz}

\newcommand{\LNO}{La$_3$Ni$_2$O$_7$}
\newcommand{\XO}{$d_{x^2-y^2}$~}
\newcommand{\ZO}{$d_{z^2}$~}

\begin{document}

\title{Orbital-selective Superconductivity in the Pressurized Bilayer Nickelate La$_3$Ni$_2$O$_7$: 
\\ An Infinite Projected Entangled-Pair State Study}

\author{Jialin Chen}
\affiliation{CAS Key Laboratory of Theoretical Physics, Institute of 
Theoretical Physics, Chinese Academy of Sciences, Beijing 100190, China}
\affiliation{Hefei National Laboratory, Hefei 230088, China} 

\author{Fan Yang}
\email{yangfan\_blg@bit.edu.cn}
\affiliation{School of Physics, Beijing Institute of Technology, Beijing 100081, China}

\author{Wei Li}
\email{w.li@itp.ac.cn}
\affiliation{CAS Key Laboratory of Theoretical Physics, Institute of Theoretical Physics, 
Chinese Academy of Sciences, Beijing 100190, China}
\affiliation{Hefei National Laboratory, Hefei 230088, China} 
 
\begin{abstract}
The newly discovered high-$T_c$ nickelate superconductor \LNO~has generated 
significant research interest. To uncover the pairing mechanism, it is essential to 
investigate the intriguing interplay between the two $e_g$, i.e., \XO and \ZO orbitals. 
Here we conduct an infinite projected entangled-pair state (iPEPS) study of the bilayer 
$t$-$J$ model, directly in the thermodynamic limit and with orbitally selective parameters 
for \XO and \ZO orbitals, respectively. The \XO electrons exhibit significant intralayer 
hopping $t_\parallel$ (and spin couplings $J_\parallel$) as well as strong interlayer 
$J_\perp$ passed from the \ZO electrons. However, the interlayer $t_\perp$ is negligible 
in this case. 
In contrast, the $d_{z^2}$ orbital demonstrates strong interlayer $t_\perp$ and $J_\perp$, 
while the inherent intralayer $t_\parallel$ and $J_\parallel$ are small. Based on the 
iPEPS results, we find clear orbital-selective behaviors in \LNO. The $d_{x^2-y^2}$ 
orbitals exhibit robust superconductive (SC) order driven by the interlayer coupling $J_\perp$; 
while the $d_{z^2}$ band shows relatively weak SC order  as a result of small $t_\parallel$ (lack of coherence) but 
large $t_\perp$ (strong Pauli blocking). Furthermore, by substituting rare-earth 
element Pm or Sm with La, we find an enhanced SC order, which opens up a promising 
avenue for discovering nickelate superconductors with even higher $T_c$.
\end{abstract}

\maketitle

\begin{figure}[!tbp]
\includegraphics[width=1\linewidth]{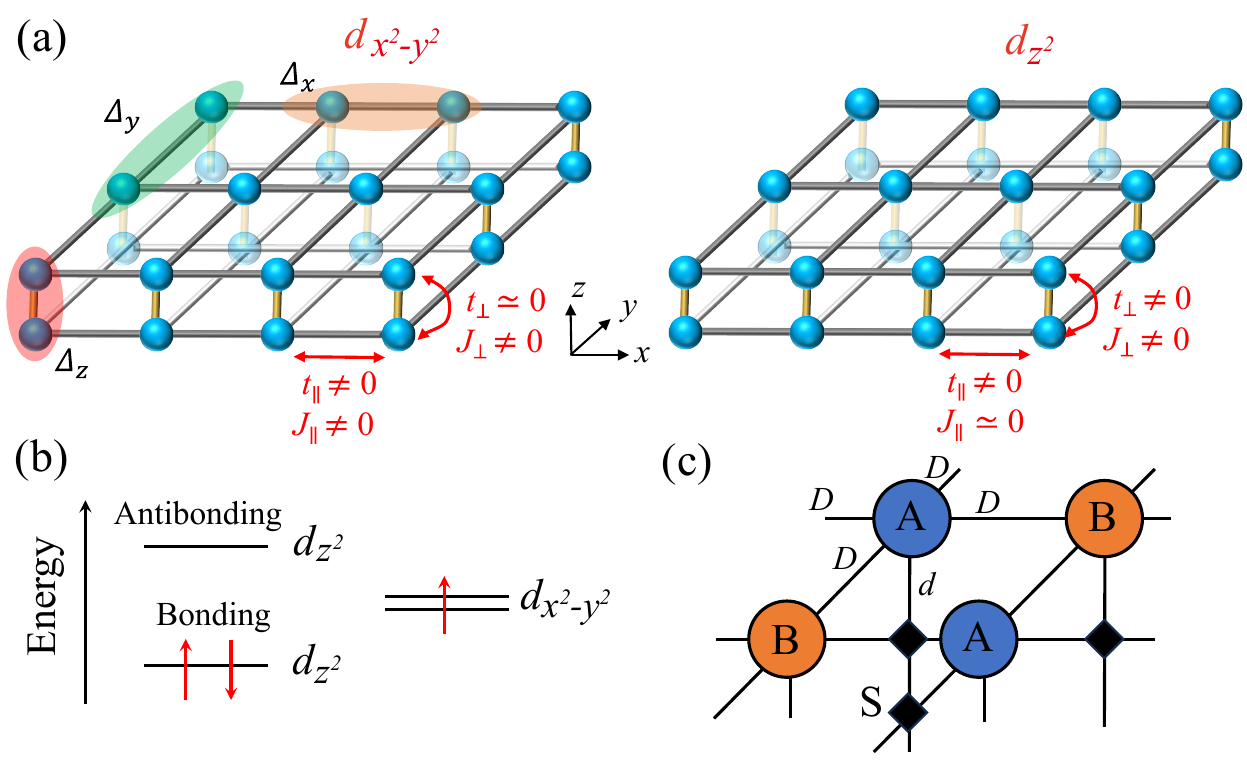} 
\caption{(a) shows the bilayer $t$-$J$ model describing the behaviors of 
$d_{x^2-y^2}$ (left) and $d_{z^2}$ (right) orbitals with properly chosen 
parameters. \XO orbital has nonzero intralayer hopping $t_\parallel$, coupling 
$J_\parallel$, and effective interlayer coupling $J_\perp$, but without interlayer 
hopping $t_\perp$. \ZO orbital has strong $t_\perp, J_\perp$ and effective 
$t_\parallel$. The SC pairing order parameters $\Delta_{x,y,z}$ on the NN 
bonds along the $x$, $y$, and $z$ axes, respectively (see definitions in the 
main text). 
(b) illustrates the energy levels for the two $e_g$ (\ZO and $d_{x^2-y^2}$) 
orbitals of the two Ni$^{2.5+}$ ($3d^{7.5}$) cations in one unit cell of the 
bilayer \LNO. 
(c) illustrates the unit cell with two different bulk tensors (A and B) used in 
the fermionic iPEPS calculations  shown in the main text. 
Swap gate $S$ is introduced to account for fermion statistics, which equals 
$-1$ when two parity-odd indices cross and 1 otherwise. $D$ and $d$ are 
the bond dimensions of the geometric and physical indices.
}
\label{Fig1}
\end{figure}

\textit{Introduction.---} 
The discovery of high-temperature superconductivity in the pressurized 
nickelate La$_3$Ni$_2$O$_7$~\cite{Nickelate80K} has raised enthusiastic 
research interest both in experiment~\cite{Liu2023correlation,Hou2023emergence,
Zhang2023hightemperature,yang2023orbitaldependent,zhang2023effects,
fwang2023pressureinduced} and theory~\cite{Luo2023Model,zhang2023electronic,
yang2023possible,lechermann2023electronic,sakakibara2023possible,
gu2023effective,shen2023effective,christiansson2023correlated,
Shilenko2023Correlated,wu2023charge,cao2023flat,chen2023critical,
liu2023spmwave,lu2023interlayer,zhang2023structural,oh2023type,
liao2023electron,qu2023bilayer,yang2023minimal,jiang2023high,
zhang2023trends,huang2023impurity,qin2023hightc,tian2023correlation,
lu2023superconductivity,jiang2023pressure,kitamine2023theoretical,
luo2023hightc,zhang2023strong,pan2023effect,sakakibara2023theoretical,
lange2023pairing,geisler2023structural,yang2023strong,rhodes2023structural,
lange2023feshbach,labollita2023electronic,kumar2023softening,kaneko2023pair,
lu2023interplay,ryee2023critical,schlomer2023superconductivity}. 
From a theoretical standpoint, the bilayer structure and orbital selectivity are two 
defining characteristics that set nickelate apart from cuprate superconductors. 
Despite significant advancements in the studies of pairing mechanisms using 
both weak and strong coupling approaches, there is still a debate regarding 
which of the two $e_g$ orbitals [c.f., Fig.~\ref{Fig1}(b)], $d_{x^2-y^2}$
\cite{lu2023interlayer,oh2023type,liao2023electron,qu2023bilayer} or 
$d_{z^2}$~\cite{yang2023minimal,qin2023hightc}, is primarily responsible 
for the robust superconductivity in La$_3$Ni$_2$O$_7$. 

Specifically, the $d_{z^2}$ orbitals have strong interlayer hopping $t_\perp$ 
and negligible intralayer hopping $t_\parallel$~\cite{Luo2023Model,
zhang2023electronic,gu2023effective}. With strong renormalization 
due to Coulomb interactions~\cite{cao2023flat,yang2023orbitaldependent}, 
the \ZO orbitals are local and have strong interlayer couplings. Thus a pair 
of electrons in the \ZO orbitals can form a localized spin-singlet dimer. 
There are theoretical proposals that suggest a pathway towards SC order, 
which involve introducing holes into the rung singlets. Hybridization with 
neighboring $e_g$ ($d_{x^2-y^2}$) orbitals provides the $d_{z^2}$ holes with 
kinetic energy~\cite{shen2023effective,yang2023minimal}. As a result, the 
tightly bound $d_{z^2}$ hole pairs can move coherently within the bilayer 
system, giving rise to long-range SC order~\cite{qin2023hightc}.

On the other hand, a contrasting viewpoint has been put forth that suggests 
the $d_{x^2-y^2}$ orbital is playing a major role in the formation of SC order 
in La$_3$Ni$_2$O$_7$~\cite{lu2023interlayer,oh2023type,qu2023bilayer,
lu2023superconductivity,pan2023effect,zhang2023strong,lange2023pairing,
lange2023feshbach,lu2023interplay}. The Hund's rule coupling with a strength 
of about 1~eV in the system~\cite{cao2023flat,christiansson2023correlated,
kumar2023softening,tian2023correlation} plays a crucial role, which transfers 
the interlayer coupling $J_\perp$ from the $d_{z^2}$ orbital to the \XO orbital 
through the symmetrization of spins on the two $e_g$ orbitals located on the 
same site. Thus a bilayer $t_{\parallel}$-$J_{\parallel}$-$J_\perp$ model well 
describes the correlated $d_{x^2-y^2}$ electrons~\cite{lu2023interlayer,
oh2023type,qu2023bilayer}, which are found to host a robust and high-$T_c$ 
SC order~\cite{lu2023interlayer,qu2023bilayer} driven by the strong 
antiferromagnetic (AFM) interlayer coupling $J_\perp$. 

In this work, we employ the fermionic infinite projected-pair state (iPEPS) 
approach, equipped with both simple (SU) and fast full updates (FFU), 
to study the bilayer $t$-$J$ model, focusing on the SC orders in the two 
$e_g$ orbitals. 
We compute the SC order parameters directly in the thermodynamic limit, 
going beyond the quasi-1D geometries in the previous density matrix 
renormalization group (DMRG) studies~\cite{qu2023bilayer,shen2023effective,
kaneko2023pair},  where only quasi-long range pairing correlations
can be obtained. Based on the accurate 2D iPEPS calculations, we find 
the \XO band can be the dominant contributor to the $s$-wave SC order 
in \LNO, while the \ZO orbital has only very weak SC pairings. 
Additionally, we explore the possibility of substituting La with other rare-earth 
elements, and find that the transition temperature $T_c$ can be enhanced 
with Pm and Sm substitutions.

\textit{Bilayer $t$-$J$ model for the \XO and \ZO orbitals.---}
There are two $e_g$ orbitals that we consider in the iPEPS calculations,
the nearly half-filled \ZO and quarter-filled \XO orbitals, each described by 
a bilayer effective model [as depicted in Fig.~\ref{Fig1}(a)],

\begin{eqnarray}
	H_{\rm bilayer} & = & -t_\parallel
	\sum_{\langle i,j \rangle, \mu, \sigma} (c^\dagger_{i, \mu, \sigma}
	c_{j, \mu, \sigma} + H.c.) \notag \\
	& + &
	J_\parallel \sum_{\langle i,j \rangle, \mu} (\bold{S}_{i,\mu} \cdot
	\bold{S}_{j,\mu} - \frac{1}{4} n_{i,\mu} n_{j,\mu}) \notag \\
	& - & t_\perp
	\sum_{i, \sigma} (c^\dagger_{i, \mu=1, \sigma}
	c_{i, \mu=-1, \sigma} + H.c.) \notag \\
	& + &
	J_{\perp} \sum_{i} \bold{S}_{i,\mu=1} \cdot
	\bold{S}_{i,\mu=-1},
	\label{Eq:b-t-J}
\end{eqnarray}
where $c^\dagger_{i, \mu, \sigma}$ ($c_{i, \mu, \sigma}$) creates 
(annihilates) an electron of spin $\sigma=\{\uparrow,\downarrow\}$ 
at site $i$ in layer $\mu=\{1,-1\}$, the vector operator $\bold{S}_{i,\mu} = 
\frac {1}{2} c^\dagger_{i,\mu,\sigma} \, (\bm{\sigma}_{\sigma,\sigma'})\, 
c_{i,\mu,\sigma'}$ denotes the spin of the electron with the Pauli matrices 
$\bm{\sigma} = \{\sigma_x,\sigma_y, \sigma_z\}$. $t_\parallel$ ($t_\perp$) 
is the intralayer (interlayer) hopping amplitude, and $J_\parallel$ 
($J_\perp$) the intralayer (interlayer) AFM coupling. The double 
occupancy is projected out in the bilayer $t$-$J$ model as usual. 

Based on the tight-binding model derived from density functional theory 
(DFT) calculations~\cite{Luo2023Model,zhang2023trends}, we choose 
$t_\parallel=1$ and $J_\parallel=1/3$ for the $d_{x^2-y^2}$ orbital, together 
with interlayer $J_\perp=2/3$ (while $t_\perp=0$) passed from the $d_{z^2}$ 
orbital~\cite{lu2023interlayer,oh2023type,qu2023bilayer}; on the other hand, 
for the $d_{z^2}$ orbital we set $t_\perp=1$ and $J_\perp=2/3$ reflecting 
the strong $\sigma$ bonding of $d_{z^2}$ electrons, with effective $t_\parallel
=1/6$ (while $J_\parallel=0$) gained from hybridization with $d_{x^2-y^2}$ 
orbitals~\cite{kaneko2023pair,shen2023effective}. We believe that the so-chosen 
parameters capture the essence of electron correlations in the two $e_g$ 
orbitals of \LNO. 

\textit{Fermionic iPEPS method.---} 
To simulate the bilayer $t$-$J$ model, we flatten the bilayer system into a 
single-layer system with enlarged local Hilbert space~\cite{qu2023bilayer} 
and employ the fermionic iPEPS method to simulate the ground state
\cite{Verstraete2004renorm,Jordan2008Classical,Cirac2021RMP,Corboz2009Fermionic,
Corboz2010Simulation,Barthel2009Contraction,Kraus2010Fermionic,
PhysRevX.8.031031,PhysRevX.8.041033,PhysRevX.8.031030}. 
As illustrated in Fig.~\ref{Fig1}(c), we set a $2\times 2$ unit cell with two 
bulk tensors A and B arranged periodically in the iPEPS wavefunction 
(larger unit cells produce consistent results, see Supplementary 
Materials~\cite{SM}), and swap gates are introduced to encode the 
fermion statistics~\cite{Corboz2009Fermionic,Corboz2010Simulation}.
Each bulk tensor has a physical bond with dimension $d=9$ representing 
the direct product of two $e_g$ orbitals with double occupancy projected out. 
The accuracy of our simulations is controlled by the geometric bond dimension 
$D$. We optimize the iPEPS wavefunction mainly using SU~\cite{Xiang2008SU,
Li2012SU,Corboz2010Simulation} with $D$ retained up to $12$ and further 
extrapolated to infinity. The FFU~\cite{FFU2015} is also exploited in the 
calculations, with bond dimension up to $D=10$, and the results 
are in great agreement with SU results~\cite{SM}. The expectation values 
are evaluated using the corner transfer matrix renormalization group method
\cite{Corboz2014Competing,Orus2009Simulation} with an environment bond 
dimension of $\chi=D^2$ that very well converge the results.
 
\begin{figure}[!tbp]
\includegraphics[width=1\linewidth]{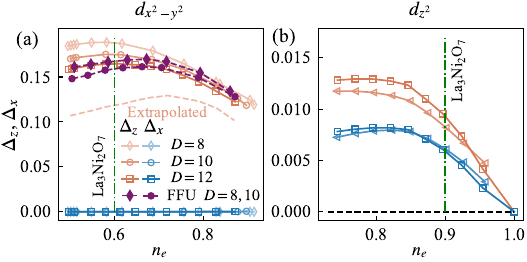}
\caption{The SC order parameters $\Delta_z$ for the interlayer pairing and 
$\Delta_x$ for the intralayer pairing, with varying electron density $n_e$ for 
(a) $d_{x^2-y^2}$ and (b) $d_{z^2}$ orbitals. $\Delta_y$ is found to be equal 
to $\Delta_x$ and thus not shown here. We retain $D$ up to 12, and for \XO 
we extrapolate $\Delta_z$ to the infinite-$D$ limit~\cite{SM}; for \ZO orbital a 
good convergence is also reached, with SC order one order of magnitude smaller 
than that of the \XO orbital. The green vertical lines mark different electron 
densities in the $d_{x^2-y^2}$ and $d_{z^2}$ orbitals, where $n_{x^2-y^2} \simeq 
0.6$ and $n_{z^2}\simeq 0.9$ in~\LNO. The model parameters are $t_\parallel=1$, 
$J_\parallel=1/3$, $t_\perp=0$, $J_\perp=2/3$ for $d_{x^2-y^2}$, and 
$t_\parallel=1/6$, $J_\parallel=0$, $t_\perp=1$, $J_\perp=2/3$ for \ZO orbital.
}
\label{Fig2} 
\end{figure}  

\begin{figure*}[!tbp]
\includegraphics[width=0.95\linewidth]{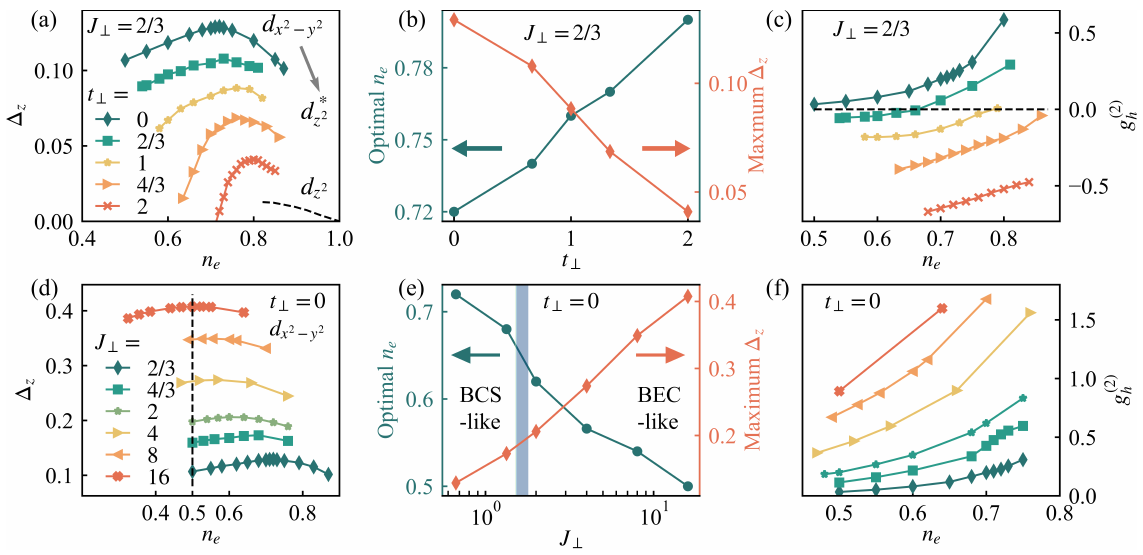}
\caption{The variation of interlayer SC order parameters $\Delta_z$ of 
$d_{x^2-y^2}$ orbital versus (a) $t_\perp$ and (d) $J_\perp$. The variations 
of maximal $\Delta_z$ and the corresponding optimal density $n_e$ are 
plotted versus $t_\perp$ and $J_\perp$ in panel (b) and (e), respectively. 
By increasing $J_\perp$ for the $d_{x^2-y^2}$ orbital, a BCS-BEC crossover 
occurs in (e). (c) and (f) show the evolution of interlayer hole correlations 
$g^{(2)}_h$ with $n_e$ for different tuning parameters, with the same legends 
as those in (a) and (d), respectively. In panel (a), we increase $t_\perp$ and 
find it changes from $d_{x^2-y^2}$ orbital-like to a coherent $d_{z^2}$ (denoted 
as $d^*_{z^2}$) behavior with weakened SC order. Besides $J_\perp$ and 
$t_\perp$ that are varying in the calculations, other model parameters are 
fixed as $t_\parallel=1$, $J_\parallel=1/3$, and all the results are extrapolated 
to infinity $D$~\cite{SM}. As a comparison, we also plot the results for the 
$d_{z^2}$ orbital taken from Fig.~\ref{Fig1}(b) with a dashed line, where the 
SC order is further reduced due to the smaller intralayer hopping $t_\parallel 
= 1/6$. The vertical dashed line in panel (d) indicates the quarter filling 
(i.e., $n=0.5$), and the shaded bar in (e) represents the BCS-BEC crossover.
}
\label{Fig3}
\end{figure*}

\textit{Orbital-selective superconductivity.---} 
In Fig.~\ref{Fig2}, we present the iPEPS results for the SC order parameters 
in the $d_{x^2-y^2}$ and $d_{z^2}$ orbitals. The $d_{x^2-y^2}$ results are 
shown in Fig.~\ref{Fig2}(a), where we compute the interlayer SC order 
parameter $\Delta_{z} = \frac{1}{\sqrt{2}} \langle \sum_{\mu=\pm1} \, 
c_{i,\mu,\uparrow}^{\dagger} c_{i,-\mu,\downarrow}^{\dagger} \rangle$ 
with SU and find a strong interlayer pairing. By increasing the electron 
density $n_e$, $\Delta_z$ first increases and then decreases, 
with a large $\Delta_z=0.13$ at the optimal density $n_e=0.72$. 
To confirm the results, in Fig.~\ref{Fig2} we also calculate $\Delta_z$ 
with FFU and find the results agree with those of SU. These mutually 
corroborative results support a robust SC order in the $d_{x^2-y^2}$ orbital.

For electron density $n_{x^2-y^2}=0.6$ relevant for the pristine compound 
\LNO~\cite{zhang2023trends, pan2023effect,shen2023effective,kaneko2023pair,
lu2023interplay}, we find the SC order parameter is $\Delta_z \simeq 0.12$,
much greater than that in a plain 2D $t$-$J$ model~\cite{Corboz2014Competing}. 
On the other hand, we find the intralayer pairings, both $\Delta_x$ and $\Delta_y$ 
[see Fig.~\ref{Fig1}(a)], are negligible for all scanned electron densities. 
Here, $\Delta_{x(y)}=\frac{1}{\sqrt{2}} \sum_{\sigma=\{\uparrow, \downarrow\}} 
\langle \rm{sgn}{(\sigma)} \, c_{i,\mu,\sigma}^{\dagger} c_{i+\hat{x}(\hat{y}), 
\mu, \bar{\sigma}}^{\dagger} \rangle$, with $\rm{sgn}(\uparrow)=1$, $\rm{sgn}
(\downarrow)=-1$, $\bar{\sigma}$ reverses the spin orientation of $\sigma$, 
and $\hat{x}$($\hat{y}$) being the unit vector whitin the square-lattice plane 
(either $\mu=1$ or $-1$).

The results for the $d_{z^2}$ orbital are presented in Fig.~\ref{Fig2}(b). 
As the electron density decreases from 1.0 to about 0.75 (i.e., hole doped), 
the magnitudes of $\Delta_z$ and $\Delta_x$ (also $\Delta_y$, not shown) 
increase and then level off for $n_e\leq0.85$ (c.f., the $D=10, 12$ data). 
The typical magnitude of $\Delta_z$ is about 0.01, one order smaller than 
that of the $d_{x^2-y^2}$ orbital shown in Fig.~\ref{Fig2}(a). These results 
indicate that the $d_{x^2-y^2}$ orbital contributes significantly more to the 
superconducting order in \LNO, consistent with recent two-orbital model 
calculations~\cite{shen2023effective,kaneko2023pair,tian2023correlation,
lu2023interplay}.

\textit{Interlayer hopping and the Pauli blocking.---} 
To understand the essential differences between the two $e_g$ orbitals 
in \LNO, we investigate the effects of the interlayer hopping $t_\perp$ 
and coupling $J_\perp$ on the SC order in Fig.~\ref{Fig3}. 
 
To study the effect of $t_\perp$, we fix $t_\parallel=1$, $J_\parallel=1/3$, 
and $J_\perp=2/3$, and tune $t_\perp$ from 0 to 2. The results are 
presented in Figs.~\ref{Fig3}(a,b), where $\Delta_z$ reduces and the 
SC dome moves towards larger density $n_e$ gradually with increasing 
$t_\perp$. We denote such coherent $d_{z^2}$ orbital as $d^*_{z^2}$, 
where we have artificially set a large $t_\parallel=1$. One possible way to 
gain such kinetic energy is through the inter-site hybridization with \XO orbital. 
Nevertheless, even for $d^*_{z^2}$ the obtained values of $\Delta_z$ are 
still significantly weakened due to the large $t_\perp$, which lead to a 
reduction in the interlayer pairing, even under the presence of strong 
interlayer coupling $J_\perp$. 

Moreover, we find that the SC order characterized by $\Delta_z$ is further 
reduced for the realistic $d_{z^2}$ orbital with smaller, but also more realistic, 
intralayer hopping $t_\parallel = 1/6$. The above two factors well explain the 
orbital-selective superconductivity observed in recent numerical calculations 
of two-orbital model~\cite{shen2023effective,tian2023correlation,kaneko2023pair}. 

To gain further insight into the effect of interlayer hopping $t_\perp$ on the SC
pairing, we study the hole-hole correlation $g_h^{(2)} \equiv \langle h_{i,\mu=1} 
h_{i,\mu=-1}  \rangle_\beta/ (\langle h_{i,\mu=1} \rangle_\beta \cdot \langle 
h_{i,\mu=-1} \rangle_\beta) - 1$, where $h_{i,\mu} = 1 - n_{i,\mu}$ counts the 
hole number. The positive (negative) values of $g_h^{(2)}$ indicate bunching 
(antibunching) of the holes. In Fig.~\ref{Fig3}(c), we observe that $g_h^{(2)}$ 
is always positive for $t_\perp = 0$, indicating occurrence of hole bunching 
between two layers. However, as $t_\perp$ increases, $g_h^{(2)}$ decreases 
and may even cross the $g_h^{(2)}=0$ line. This is because the interlayer 
hopping $t_\perp$ can introduce statistical repulsion between holes and is 
detrimental to interlayer pairing~\cite{Hilker2023pairing}. The electron density 
at the point where $g_h^{(2)}$ crosses zero gradually increases with increasing 
$t_\perp$ in Fig.~\ref{Fig3}(c), consistent with the observation that the SC dome 
moves towards larger $n_e$ values as $t_\perp$ increases in Fig.~\ref{Fig3}(a).

\textit{Interlayer coupling driven BCS-BEC crossover.---}
In the $d_{x^2-y^2}$ orbital scenario, the interlayer $J_\perp$ plays an 
essential role in driving the SC pairing. To reveal the advantage and explore 
the limit of the SC order in the $d_{x^2-y^2}$ orbital, in Fig.~\ref{Fig3}(d-f) 
we present the results computed with model parameters $t_\parallel=1$, 
$J_\parallel=1/3$, and $t_\perp=0$, similar to those used in Fig.~\ref{Fig2}(a), 
but with an increased AFM coupling $J_\perp$. In Fig.~\ref{Fig3}(d) we find 
that as $J_\perp$ increases, the interlayer SC order $\Delta_z$ increases 
and the SC dome shifts towards smaller $n_e$. To show the effect of 
$J_\perp$ more clearly, we collect the data and plot $\Delta_z$ versus 
$J_\perp$ in Fig.~\ref{Fig3}(e), and observe that the maximum $\Delta_z$ 
increases drastically from about 0.13 to 0.41. The optimal $n_e$ decreases 
from 0.72 to 0.5 (i.e., quarter filling), in agreement with recent analytical 
results on the $t_\parallel$-$J_\parallel$-$J_\perp$ model
\cite{lu2023superconductivity,zhang2023strong}. 

The strong interlayer pairing in $d_{x^2-y^2}$ orbital can also be witnessed 
by the positive $g_h^{(2)}$ shown in Fig.~\ref{Fig3}(f), which represents a 
strong bunching between the two holes on the same interlayer vertical bond. 
We find that $g_h^{(2)}$ is always positive and the hole bunching becomes 
greater as $J_\perp$ increases. For sufficiently large $J_\perp$, the hole pair 
changes from a loosely bounded Cooper pair as in the Bardeen-Cooper-Schrieffer 
(BCS) theory, to a tightly bounded pair like a boson in the Bose-Einstein 
condensation (BEC). The maximal $\Delta_z$ appears at electron density $n=0.5$, 
where the bosons gain the highest mobility. Therefore, the evolution of optimal 
density $n_e$ from $0.72$ to $0.5$ indicates that a BCS-BEC crossover by
increasing $J_\perp$~\citep{lu2023superconductivity}, and the realistic value 
$J_\perp/t_\parallel \approx 2/3$ places the compound \LNO~in the BCS side. 
These results highlight the potential of compounds with a similar bilayer structure
to \LNO~as a highly promising family of superconductors, with the possibility of 
achieving even higher $T_c$.

\textit{Mixed-dimensional bilayer pairing in \LNO.---}
In addition to the absence of coherent behavior and small hole densities that 
are essential in preventing the \ZO orbital from achieving robust high-$T_c$ 
superconductivity~\cite{lu2023interlayer,lu2023interplay}, we emphasize that 
the mixD bilayer structure is another critical factor that distinguishes the two 
$e_g$ orbitals.

Specifically, for the \ZO orbital the optimal electron density is close to 
half-filling, i.e., $\gtrsim 0.8$, similar to conventional single-layer Hubbard 
or $t$-$J$ system~\cite{Corboz2014Competing}. 
On the other hand, the \XO orbital can be regarded to realize a mixD bilayer 
system~\cite{Grusdt2022mixD,Hilker2023pairing}, which has inter- and 
intralayer spin couplings ($J_{\perp}, J_{\parallel}$) as well as intralayer 
hopping $t_\parallel$ but no interlayer hopping $t_\perp$. Such a mixD 
bilayer system benefits from a strong pairing force arising from the large
AFM coupling $J_\perp$ and avoids the Pauli blocking due to the absence 
of interlayer $t_\perp$. As a result, the \XO orbital with the mixD bilayer 
structure is dominating in forming the SC order, which becomes progressively 
weakened as one approaches the more conventional bilayer structure of 
$d^*_{z^2}$ orbitals by increasing $t_\perp$ [see Fig.~\ref{Fig3}(a)]. 
    
\begin{figure}[!tbp] 
\includegraphics[width=1.0\linewidth]{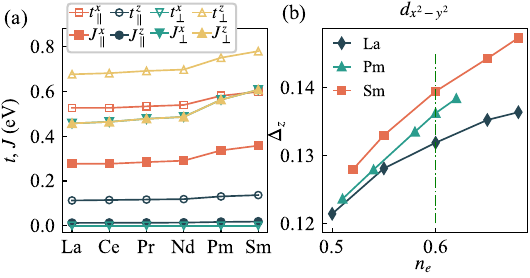}
\caption{(a) Hopping amplitudes and AFM couplings for the element substituted 
R$_3$Ni$_2$O$_7$ with R from La to Sm, and the superscript $x$ ($z$) 
represents the $d_{x^2-y^2}$ ($d_{z^2}$) orbital. In the strong Hund's coupling 
limit, the interlayer AFM coupling can be fully passed from \ZO orbital to 
the \XO one, namely, $J^x_{\perp}\equiv J^z_{\perp}$~\cite{lu2023interlayer,
qu2023bilayer}. (b) The computed SC order parameter $\Delta_z$ versus 
density $n_e$ for the $d_{x^2-y^2}$ orbital, with R = La, Pm, and Sm. 
The green vertical line marks the estimated electron densities $n_e=0.6$ 
for R$_3$Ni$_2$O$_7$. All SU results shown have been extrapolating to 
infinite $D$~\cite{SM}.}
\label{Fig4}
\end{figure}

\textit{Enhanced SC in R$_3$Ni$_2$O$_7$ with element substitution.---} 
Recently, DFT calculations showed that the Fmmm crystal structure is 
retained under pressure for rare-earth (RE) element substitution
\cite{zhang2023trends}, where the hopping amplitudes and also exchange 
interactions can be enhanced [c.f., Fig.~\ref{Fig4}(a)]. The authors in Ref.
\cite{zhang2023trends} further predicted that the pairing and $T_c$ would 
decrease with such RE substitution from La to Sm, and that \LNO~is already 
``optimal". On the other hand, in Ref.~\cite{pan2023effect}, a strong-coupling 
analysis based on slave boson mean-field theory predicted that the RE 
substitution can significantly enhance the pairing strength and thus $T_c$, 
in sharp contrast to the weak coupling analysis~\cite{zhang2023trends}.

To settle this debate, we carry out iPEPS calculations with realistic 
parameters obtained from the DFT calculations~\cite{zhang2023trends} 
shown in Fig.~\ref{Fig4}(a). With properly chosen Coulomb interaction 
$U=4$~eV~\cite{zhang2023trends,pan2023effect,yang2023orbitaldependent}, 
we estimate the AFM exchange interactions $J^z_{\perp}$ and $J^z_\parallel$ 
for the $d_{z^2}$ orbital and $J^x_\parallel$ for the $d_{x^2-y^2}$ orbital 
according to the superexchange $J=4t^2/U$. As shown in Fig.~\ref{Fig4}(b), 
the obtained SC order parameter $\Delta_z$ of the $d_{x^2-y^2}$ orbital 
increases when substituting La from Pm to Sm, at density $n_e=0.6$ 
relevant for the nickelates. These results support that the SC pairing can 
be strengthened by element substitution, in agreement with the conclusion in 
Ref.~\cite{pan2023effect} from the strong-coupling approach. By inspecting 
the hopping and coupling parameters in Fig.~\ref{Fig4}(a), we find the 
enhancement of SC order mainly originates from the increased interlayer 
AFM interactions after the element substitution.

\textit{Discussion and outlook.---}
In this work, we perform iPEPS simulations of the single-orbital bilayer 
$t$-$J$ model for \XO or \ZO orbital in La$_3$Ni$_2$O$_7$, directly 
in the thermodynamic limit, with corroborative simple and full update 
optimizations. Our results indicate that the interlayer superconducting order 
in the $d_{x^2-y^2}$ orbital is significantly stronger compared to that in the 
$d_{z^2}$ orbital, due to the mixD bilayer structure that facilitates the SC 
order. The orbital selectivity originates from the different values of $t_\perp$ and 
$t_\parallel$ in the two orbitals, which have distinct effects on the SC 
order. $t_\perp$ can introduce Pauli blocking that is destructive 
for interlayer pairing, while a sufficiently large $t_\parallel$ is needed to 
render phase coherence for long-range SC order.

Our findings highlight the intriguing connections between two seemingly separate 
fields: the high-$T_c$ nickelate superconductors and the optical lattice quantum 
simulations. In the latter, the mixD ladder system has been realized
\cite{Hilker2023pairing} and intensively discussed~\cite{lange2023pairing,
yang2023strong,lange2023feshbach} recently. One possible extension of 
the present study is to include the $T>0$ tensor-network calculations
\cite{Li2011a,Dong2017,tanTRG2023,White2009METTS,Stoudenmire2010,
Czarnik2014PEPO,Czarnik2016PRB} relevant for the nickelate and quantum 
gas experiments. 

Lastly, while our comparative study of the \XO and \ZO orbitals provide 
insights into the orbital-selective behaviors, a comprehensive two-orbital 
bilayer $t$-$J$ model that includes both $e_g$ orbitals is necessary to 
fully address their roles in \LNO. There were attempts to study this 
interplay with DMRG calculations in ladder systems~\cite{shen2023effective,
kaneko2023pair}. However, the study of two coupled infinite layers 
still poses significant challenges and is left for future studies.

\begin{acknowledgments}
\textit{Acknowledgments.---} JC and WL are indebted to Xing-Zhou Qu, 
Dai-Wei Qu, Xing-Yu Zhang, Lei Wang, and Gang Su for stimulating discussions. 
This work was supported by the National Natural Science Foundation of 
China (Grant Nos.~12222412, 11974036, 12047503), Innovation Program 
for Quantum Science and Technology (Nos.~2021ZD0301900), and CAS 
Project for Young Scientists in Basic Research (Grant No.~YSBR-057). 
We thank the HPC-ITP for the technical support and generous allocation 
of CPU time.
\end{acknowledgments}

\bibliography{nickelate.bib}

\begin{thebibliography}{74}%
\makeatletter
\providecommand \@ifxundefined [1]{%
 \@ifx{#1\undefined}
}%
\providecommand \@ifnum [1]{%
 \ifnum #1\expandafter \@firstoftwo
 \else \expandafter \@secondoftwo
 \fi
}%
\providecommand \@ifx [1]{%
 \ifx #1\expandafter \@firstoftwo
 \else \expandafter \@secondoftwo
 \fi
}%
\providecommand \natexlab [1]{#1}%
\providecommand \enquote  [1]{``#1''}%
\providecommand \bibnamefont  [1]{#1}%
\providecommand \bibfnamefont [1]{#1}%
\providecommand \citenamefont [1]{#1}%
\providecommand \href@noop [0]{\@secondoftwo}%
\providecommand \href [0]{\begingroup \@sanitize@url \@href}%
\providecommand \@href[1]{\@@startlink{#1}\@@href}%
\providecommand \@@href[1]{\endgroup#1\@@endlink}%
\providecommand \@sanitize@url [0]{\catcode `\\12\catcode `\$12\catcode
  `\&12\catcode `\#12\catcode `\^12\catcode `\_12\catcode `\%12\relax}%
\providecommand \@@startlink[1]{}%
\providecommand \@@endlink[0]{}%
\providecommand \url  [0]{\begingroup\@sanitize@url \@url }%
\providecommand \@url [1]{\endgroup\@href {#1}{\urlprefix }}%
\providecommand \urlprefix  [0]{URL }%
\providecommand \Eprint [0]{\href }%
\providecommand \doibase [0]{https://doi.org/}%
\providecommand \selectlanguage [0]{\@gobble}%
\providecommand \bibinfo  [0]{\@secondoftwo}%
\providecommand \bibfield  [0]{\@secondoftwo}%
\providecommand \translation [1]{[#1]}%
\providecommand \BibitemOpen [0]{}%
\providecommand \bibitemStop [0]{}%
\providecommand \bibitemNoStop [0]{.\EOS\space}%
\providecommand \EOS [0]{\spacefactor3000\relax}%
\providecommand \BibitemShut  [1]{\csname bibitem#1\endcsname}%
\let\auto@bib@innerbib\@empty
\bibitem [{\citenamefont {Sun}\ \emph {et~al.}(2023)\citenamefont {Sun},
  \citenamefont {Huo}, \citenamefont {Hu}, \citenamefont {Li}, \citenamefont
  {Liu}, \citenamefont {Han}, \citenamefont {Tang}, \citenamefont {Mao},
  \citenamefont {Yang}, \citenamefont {Wang}, \citenamefont {Cheng},
  \citenamefont {Yao}, \citenamefont {Zhang},\ and\ \citenamefont
  {Wang}}]{Nickelate80K}%
  \BibitemOpen
  \bibfield  {author} {\bibinfo {author} {\bibfnamefont {H.}~\bibnamefont
  {Sun}}, \bibinfo {author} {\bibfnamefont {M.}~\bibnamefont {Huo}}, \bibinfo
  {author} {\bibfnamefont {X.}~\bibnamefont {Hu}}, \bibinfo {author}
  {\bibfnamefont {J.}~\bibnamefont {Li}}, \bibinfo {author} {\bibfnamefont
  {Z.}~\bibnamefont {Liu}}, \bibinfo {author} {\bibfnamefont {Y.}~\bibnamefont
  {Han}}, \bibinfo {author} {\bibfnamefont {L.}~\bibnamefont {Tang}}, \bibinfo
  {author} {\bibfnamefont {Z.}~\bibnamefont {Mao}}, \bibinfo {author}
  {\bibfnamefont {P.}~\bibnamefont {Yang}}, \bibinfo {author} {\bibfnamefont
  {B.}~\bibnamefont {Wang}}, \bibinfo {author} {\bibfnamefont {J.}~\bibnamefont
  {Cheng}}, \bibinfo {author} {\bibfnamefont {D.-X.}\ \bibnamefont {Yao}},
  \bibinfo {author} {\bibfnamefont {G.-M.}\ \bibnamefont {Zhang}},\ and\
  \bibinfo {author} {\bibfnamefont {M.}~\bibnamefont {Wang}},\ }\bibfield
  {title} {\bibinfo {title} {Signatures of superconductivity near {80 K} in a
  nickelate under high pressure},\ }\href
  {https://doi.org/10.1038/s41586-023-06408-7} {\bibfield  {journal} {\bibinfo
  {journal} {Nature}\ }\textbf {\bibinfo {volume} {621}},\ \bibinfo {pages}
  {493} (\bibinfo {year} {2023})}\BibitemShut {NoStop}%
\bibitem [{\citenamefont {Liu}\ \emph {et~al.}(2023{\natexlab{a}})\citenamefont
  {Liu}, \citenamefont {Huo}, \citenamefont {Li}, \citenamefont {Li},
  \citenamefont {Liu}, \citenamefont {Dai}, \citenamefont {Zhou}, \citenamefont
  {Hao}, \citenamefont {Lu}, \citenamefont {Wang},\ and\ \citenamefont
  {Wen}}]{Liu2023correlation}%
  \BibitemOpen
  \bibfield  {author} {\bibinfo {author} {\bibfnamefont {Z.}~\bibnamefont
  {Liu}}, \bibinfo {author} {\bibfnamefont {M.}~\bibnamefont {Huo}}, \bibinfo
  {author} {\bibfnamefont {J.}~\bibnamefont {Li}}, \bibinfo {author}
  {\bibfnamefont {Q.}~\bibnamefont {Li}}, \bibinfo {author} {\bibfnamefont
  {Y.}~\bibnamefont {Liu}}, \bibinfo {author} {\bibfnamefont {Y.}~\bibnamefont
  {Dai}}, \bibinfo {author} {\bibfnamefont {X.}~\bibnamefont {Zhou}}, \bibinfo
  {author} {\bibfnamefont {J.}~\bibnamefont {Hao}}, \bibinfo {author}
  {\bibfnamefont {Y.}~\bibnamefont {Lu}}, \bibinfo {author} {\bibfnamefont
  {M.}~\bibnamefont {Wang}},\ and\ \bibinfo {author} {\bibfnamefont {H.-H.}\
  \bibnamefont {Wen}},\ }\href@noop {} {\bibinfo {title} {Electronic
  correlations and energy gap in the bilayer nickelate {La$_3$Ni$_2$O$_7$}}}
  (\bibinfo {year} {2023}{\natexlab{a}}),\ \Eprint
  {https://arxiv.org/abs/2307.02950} {arXiv:2307.02950 [cond-mat.supr-con]}
  \BibitemShut {NoStop}%
\bibitem [{\citenamefont {Hou}\ \emph {et~al.}(2023)\citenamefont {Hou},
  \citenamefont {Yang}, \citenamefont {Liu}, \citenamefont {Li}, \citenamefont
  {Shan}, \citenamefont {Ma}, \citenamefont {Wang}, \citenamefont {Wang},
  \citenamefont {Guo}, \citenamefont {Sun}, \citenamefont {Uwatoko},
  \citenamefont {Wang}, \citenamefont {Zhang}, \citenamefont {Wang},\ and\
  \citenamefont {Cheng}}]{Hou2023emergence}%
  \BibitemOpen
  \bibfield  {author} {\bibinfo {author} {\bibfnamefont {J.}~\bibnamefont
  {Hou}}, \bibinfo {author} {\bibfnamefont {P.-T.}\ \bibnamefont {Yang}},
  \bibinfo {author} {\bibfnamefont {Z.-Y.}\ \bibnamefont {Liu}}, \bibinfo
  {author} {\bibfnamefont {J.-Y.}\ \bibnamefont {Li}}, \bibinfo {author}
  {\bibfnamefont {P.-F.}\ \bibnamefont {Shan}}, \bibinfo {author}
  {\bibfnamefont {L.}~\bibnamefont {Ma}}, \bibinfo {author} {\bibfnamefont
  {G.}~\bibnamefont {Wang}}, \bibinfo {author} {\bibfnamefont {N.-N.}\
  \bibnamefont {Wang}}, \bibinfo {author} {\bibfnamefont {H.-Z.}\ \bibnamefont
  {Guo}}, \bibinfo {author} {\bibfnamefont {J.-P.}\ \bibnamefont {Sun}},
  \bibinfo {author} {\bibfnamefont {Y.}~\bibnamefont {Uwatoko}}, \bibinfo
  {author} {\bibfnamefont {M.}~\bibnamefont {Wang}}, \bibinfo {author}
  {\bibfnamefont {G.-M.}\ \bibnamefont {Zhang}}, \bibinfo {author}
  {\bibfnamefont {B.-S.}\ \bibnamefont {Wang}},\ and\ \bibinfo {author}
  {\bibfnamefont {J.-G.}\ \bibnamefont {Cheng}},\ }\bibfield  {title} {\bibinfo
  {title} {Emergence of high-temperature superconducting phase in pressurized
  {La$_{3}$Ni$_{2}$O$_7$} crystals},\ }\href
  {https://doi.org/10.1088/0256-307X/40/11/117302} {\bibfield  {journal}
  {\bibinfo  {journal} {Chinese Physics Letters}\ }\textbf {\bibinfo {volume}
  {40}},\ \bibinfo {eid} {117302} (\bibinfo {year} {2023})}\BibitemShut
  {NoStop}%
\bibitem [{\citenamefont {Zhang}\ \emph
  {et~al.}(2023{\natexlab{a}})\citenamefont {Zhang}, \citenamefont {Su},
  \citenamefont {Huang}, \citenamefont {Sun}, \citenamefont {Huo},
  \citenamefont {Shan}, \citenamefont {Ye}, \citenamefont {Yang}, \citenamefont
  {Li}, \citenamefont {Smidman}, \citenamefont {Wang}, \citenamefont {Jiao},\
  and\ \citenamefont {Yuan}}]{Zhang2023hightemperature}%
  \BibitemOpen
  \bibfield  {author} {\bibinfo {author} {\bibfnamefont {Y.}~\bibnamefont
  {Zhang}}, \bibinfo {author} {\bibfnamefont {D.}~\bibnamefont {Su}}, \bibinfo
  {author} {\bibfnamefont {Y.}~\bibnamefont {Huang}}, \bibinfo {author}
  {\bibfnamefont {H.}~\bibnamefont {Sun}}, \bibinfo {author} {\bibfnamefont
  {M.}~\bibnamefont {Huo}}, \bibinfo {author} {\bibfnamefont {Z.}~\bibnamefont
  {Shan}}, \bibinfo {author} {\bibfnamefont {K.}~\bibnamefont {Ye}}, \bibinfo
  {author} {\bibfnamefont {Z.}~\bibnamefont {Yang}}, \bibinfo {author}
  {\bibfnamefont {R.}~\bibnamefont {Li}}, \bibinfo {author} {\bibfnamefont
  {M.}~\bibnamefont {Smidman}}, \bibinfo {author} {\bibfnamefont
  {M.}~\bibnamefont {Wang}}, \bibinfo {author} {\bibfnamefont {L.}~\bibnamefont
  {Jiao}},\ and\ \bibinfo {author} {\bibfnamefont {H.}~\bibnamefont {Yuan}},\
  }\href@noop {} {\bibinfo {title} {High-temperature superconductivity with
  zero-resistance and strange metal behavior in {La$_3$Ni$_2$O$_7$}}} (\bibinfo
  {year} {2023}{\natexlab{a}}),\ \Eprint {https://arxiv.org/abs/2307.14819}
  {arXiv:2307.14819 [cond-mat.supr-con]} \BibitemShut {NoStop}%
\bibitem [{\citenamefont {Yang}\ \emph
  {et~al.}(2023{\natexlab{a}})\citenamefont {Yang}, \citenamefont {Sun},
  \citenamefont {Hu}, \citenamefont {Xie}, \citenamefont {Miao}, \citenamefont
  {Luo}, \citenamefont {Chen}, \citenamefont {Liang}, \citenamefont {Zhu},
  \citenamefont {Qu}, \citenamefont {Chen}, \citenamefont {Huo}, \citenamefont
  {Huang}, \citenamefont {Zhang}, \citenamefont {Zhang}, \citenamefont {Yang},
  \citenamefont {Wang}, \citenamefont {Peng}, \citenamefont {Mao},
  \citenamefont {Liu}, \citenamefont {Xu}, \citenamefont {Qian}, \citenamefont
  {Yao}, \citenamefont {Wang}, \citenamefont {Zhao},\ and\ \citenamefont
  {Zhou}}]{yang2023orbitaldependent}%
  \BibitemOpen
  \bibfield  {author} {\bibinfo {author} {\bibfnamefont {J.}~\bibnamefont
  {Yang}}, \bibinfo {author} {\bibfnamefont {H.}~\bibnamefont {Sun}}, \bibinfo
  {author} {\bibfnamefont {X.}~\bibnamefont {Hu}}, \bibinfo {author}
  {\bibfnamefont {Y.}~\bibnamefont {Xie}}, \bibinfo {author} {\bibfnamefont
  {T.}~\bibnamefont {Miao}}, \bibinfo {author} {\bibfnamefont {H.}~\bibnamefont
  {Luo}}, \bibinfo {author} {\bibfnamefont {H.}~\bibnamefont {Chen}}, \bibinfo
  {author} {\bibfnamefont {B.}~\bibnamefont {Liang}}, \bibinfo {author}
  {\bibfnamefont {W.}~\bibnamefont {Zhu}}, \bibinfo {author} {\bibfnamefont
  {G.}~\bibnamefont {Qu}}, \bibinfo {author} {\bibfnamefont {C.-Q.}\
  \bibnamefont {Chen}}, \bibinfo {author} {\bibfnamefont {M.}~\bibnamefont
  {Huo}}, \bibinfo {author} {\bibfnamefont {Y.}~\bibnamefont {Huang}}, \bibinfo
  {author} {\bibfnamefont {S.}~\bibnamefont {Zhang}}, \bibinfo {author}
  {\bibfnamefont {F.}~\bibnamefont {Zhang}}, \bibinfo {author} {\bibfnamefont
  {F.}~\bibnamefont {Yang}}, \bibinfo {author} {\bibfnamefont {Z.}~\bibnamefont
  {Wang}}, \bibinfo {author} {\bibfnamefont {Q.}~\bibnamefont {Peng}}, \bibinfo
  {author} {\bibfnamefont {H.}~\bibnamefont {Mao}}, \bibinfo {author}
  {\bibfnamefont {G.}~\bibnamefont {Liu}}, \bibinfo {author} {\bibfnamefont
  {Z.}~\bibnamefont {Xu}}, \bibinfo {author} {\bibfnamefont {T.}~\bibnamefont
  {Qian}}, \bibinfo {author} {\bibfnamefont {D.-X.}\ \bibnamefont {Yao}},
  \bibinfo {author} {\bibfnamefont {M.}~\bibnamefont {Wang}}, \bibinfo {author}
  {\bibfnamefont {L.}~\bibnamefont {Zhao}},\ and\ \bibinfo {author}
  {\bibfnamefont {X.~J.}\ \bibnamefont {Zhou}},\ }\href@noop {} {\bibinfo
  {title} {Orbital-dependent electron correlation in double-layer nickelate
  {La$_3$Ni$_2$O$_7$}}} (\bibinfo {year} {2023}{\natexlab{a}}),\ \Eprint
  {https://arxiv.org/abs/2309.01148} {arXiv:2309.01148 [cond-mat.supr-con]}
  \BibitemShut {NoStop}%
\bibitem [{\citenamefont {Zhang}\ \emph
  {et~al.}(2024{\natexlab{a}})\citenamefont {Zhang}, \citenamefont {Pei},
  \citenamefont {Wang}, \citenamefont {Zhao}, \citenamefont {Li}, \citenamefont
  {Cao}, \citenamefont {Zhu}, \citenamefont {Wu},\ and\ \citenamefont
  {Qi}}]{zhang2023effects}%
  \BibitemOpen
  \bibfield  {author} {\bibinfo {author} {\bibfnamefont {M.}~\bibnamefont
  {Zhang}}, \bibinfo {author} {\bibfnamefont {C.}~\bibnamefont {Pei}}, \bibinfo
  {author} {\bibfnamefont {Q.}~\bibnamefont {Wang}}, \bibinfo {author}
  {\bibfnamefont {Y.}~\bibnamefont {Zhao}}, \bibinfo {author} {\bibfnamefont
  {C.}~\bibnamefont {Li}}, \bibinfo {author} {\bibfnamefont {W.}~\bibnamefont
  {Cao}}, \bibinfo {author} {\bibfnamefont {S.}~\bibnamefont {Zhu}}, \bibinfo
  {author} {\bibfnamefont {J.}~\bibnamefont {Wu}},\ and\ \bibinfo {author}
  {\bibfnamefont {Y.}~\bibnamefont {Qi}},\ }\bibfield  {title} {\bibinfo
  {title} {Effects of pressure and doping on {Ruddlesden-Popper} phases
  {La$_{n+1}$Ni$_n$O$_{3n+1}$}},\ }\href
  {https://doi.org/https://doi.org/10.1016/j.jmst.2023.11.011} {\bibfield
  {journal} {\bibinfo  {journal} {Journal of Materials Science \& Technology}\
  }\textbf {\bibinfo {volume} {185}},\ \bibinfo {pages} {147} (\bibinfo {year}
  {2024}{\natexlab{a}})}\BibitemShut {NoStop}%
\bibitem [{\citenamefont {Wang}\ \emph {et~al.}(2024)\citenamefont {Wang},
  \citenamefont {Wang}, \citenamefont {Shen}, \citenamefont {Hou},
  \citenamefont {Ma}, \citenamefont {Shi}, \citenamefont {Ren}, \citenamefont
  {Gu}, \citenamefont {Ma}, \citenamefont {Yang}, \citenamefont {Liu},
  \citenamefont {Guo}, \citenamefont {Sun}, \citenamefont {Zhang},
  \citenamefont {Calder}, \citenamefont {Yan}, \citenamefont {Wang},
  \citenamefont {Uwatoko},\ and\ \citenamefont
  {Cheng}}]{fwang2023pressureinduced}%
  \BibitemOpen
  \bibfield  {author} {\bibinfo {author} {\bibfnamefont {G.}~\bibnamefont
  {Wang}}, \bibinfo {author} {\bibfnamefont {N.~N.}\ \bibnamefont {Wang}},
  \bibinfo {author} {\bibfnamefont {X.~L.}\ \bibnamefont {Shen}}, \bibinfo
  {author} {\bibfnamefont {J.}~\bibnamefont {Hou}}, \bibinfo {author}
  {\bibfnamefont {L.}~\bibnamefont {Ma}}, \bibinfo {author} {\bibfnamefont
  {L.~F.}\ \bibnamefont {Shi}}, \bibinfo {author} {\bibfnamefont {Z.~A.}\
  \bibnamefont {Ren}}, \bibinfo {author} {\bibfnamefont {Y.~D.}\ \bibnamefont
  {Gu}}, \bibinfo {author} {\bibfnamefont {H.~M.}\ \bibnamefont {Ma}}, \bibinfo
  {author} {\bibfnamefont {P.~T.}\ \bibnamefont {Yang}}, \bibinfo {author}
  {\bibfnamefont {Z.~Y.}\ \bibnamefont {Liu}}, \bibinfo {author} {\bibfnamefont
  {H.~Z.}\ \bibnamefont {Guo}}, \bibinfo {author} {\bibfnamefont {J.~P.}\
  \bibnamefont {Sun}}, \bibinfo {author} {\bibfnamefont {G.~M.}\ \bibnamefont
  {Zhang}}, \bibinfo {author} {\bibfnamefont {S.}~\bibnamefont {Calder}},
  \bibinfo {author} {\bibfnamefont {J.-Q.}\ \bibnamefont {Yan}}, \bibinfo
  {author} {\bibfnamefont {B.~S.}\ \bibnamefont {Wang}}, \bibinfo {author}
  {\bibfnamefont {Y.}~\bibnamefont {Uwatoko}},\ and\ \bibinfo {author}
  {\bibfnamefont {J.-G.}\ \bibnamefont {Cheng}},\ }\bibfield  {title} {\bibinfo
  {title} {Pressure-induced superconductivity in polycrystalline
  {${\mathrm{La}}_{3}{\mathrm{Ni}}_{2}{\mathrm{O}}_{7\ensuremath{-}\ensuremath{\delta}}$}},\
  }\href {https://doi.org/10.1103/PhysRevX.14.011040} {\bibfield  {journal}
  {\bibinfo  {journal} {Phys. Rev. X}\ }\textbf {\bibinfo {volume} {14}},\
  \bibinfo {pages} {011040} (\bibinfo {year} {2024})}\BibitemShut {NoStop}%
\bibitem [{\citenamefont {Luo}\ \emph {et~al.}(2023{\natexlab{a}})\citenamefont
  {Luo}, \citenamefont {Hu}, \citenamefont {Wang}, \citenamefont {W\'u},\ and\
  \citenamefont {Yao}}]{Luo2023Model}%
  \BibitemOpen
  \bibfield  {author} {\bibinfo {author} {\bibfnamefont {Z.}~\bibnamefont
  {Luo}}, \bibinfo {author} {\bibfnamefont {X.}~\bibnamefont {Hu}}, \bibinfo
  {author} {\bibfnamefont {M.}~\bibnamefont {Wang}}, \bibinfo {author}
  {\bibfnamefont {W.}~\bibnamefont {W\'u}},\ and\ \bibinfo {author}
  {\bibfnamefont {D.-X.}\ \bibnamefont {Yao}},\ }\bibfield  {title} {\bibinfo
  {title} {Bilayer two-orbital model of
  {$\mathrm{L}{\mathrm{a}}_{3}\mathrm{N}{\mathrm{i}}_{2}{\mathrm{O}}_{7}$}
  under pressure},\ }\href {https://doi.org/10.1103/PhysRevLett.131.126001}
  {\bibfield  {journal} {\bibinfo  {journal} {Phys. Rev. Lett.}\ }\textbf
  {\bibinfo {volume} {131}},\ \bibinfo {pages} {126001} (\bibinfo {year}
  {2023}{\natexlab{a}})}\BibitemShut {NoStop}%
\bibitem [{\citenamefont {Zhang}\ \emph
  {et~al.}(2023{\natexlab{b}})\citenamefont {Zhang}, \citenamefont {Lin},
  \citenamefont {Moreo},\ and\ \citenamefont {Dagotto}}]{zhang2023electronic}%
  \BibitemOpen
  \bibfield  {author} {\bibinfo {author} {\bibfnamefont {Y.}~\bibnamefont
  {Zhang}}, \bibinfo {author} {\bibfnamefont {L.-F.}\ \bibnamefont {Lin}},
  \bibinfo {author} {\bibfnamefont {A.}~\bibnamefont {Moreo}},\ and\ \bibinfo
  {author} {\bibfnamefont {E.}~\bibnamefont {Dagotto}},\ }\bibfield  {title}
  {\bibinfo {title} {{Electronic structure, dimer physics, orbital-selective
  behavior, and magnetic tendencies in the bilayer nickelate superconductor
  ${\mathrm{La}}_{3}{\mathrm{Ni}}_{2}{\mathrm{O}}_{7}$ under pressure}},\
  }\href {https://doi.org/10.1103/PhysRevB.108.L180510} {\bibfield  {journal}
  {\bibinfo  {journal} {Phys. Rev. B}\ }\textbf {\bibinfo {volume} {108}},\
  \bibinfo {pages} {L180510} (\bibinfo {year}
  {2023}{\natexlab{b}})}\BibitemShut {NoStop}%
\bibitem [{\citenamefont {Yang}\ \emph
  {et~al.}(2023{\natexlab{b}})\citenamefont {Yang}, \citenamefont {Wang},\ and\
  \citenamefont {Wang}}]{yang2023possible}%
  \BibitemOpen
  \bibfield  {author} {\bibinfo {author} {\bibfnamefont {Q.-G.}\ \bibnamefont
  {Yang}}, \bibinfo {author} {\bibfnamefont {D.}~\bibnamefont {Wang}},\ and\
  \bibinfo {author} {\bibfnamefont {Q.-H.}\ \bibnamefont {Wang}},\ }\bibfield
  {title} {\bibinfo {title} {Possible ${s}_{\ifmmode\pm\else\textpm\fi{}}$-wave
  superconductivity in
  {${\mathrm{La}}_{3}{\mathrm{Ni}}_{2}{\mathrm{O}}_{7}$}},\ }\href
  {https://doi.org/10.1103/PhysRevB.108.L140505} {\bibfield  {journal}
  {\bibinfo  {journal} {Phys. Rev. B}\ }\textbf {\bibinfo {volume} {108}},\
  \bibinfo {pages} {L140505} (\bibinfo {year}
  {2023}{\natexlab{b}})}\BibitemShut {NoStop}%
\bibitem [{\citenamefont {Lechermann}\ \emph {et~al.}(2023)\citenamefont
  {Lechermann}, \citenamefont {Gondolf}, \citenamefont {B\"otzel},\ and\
  \citenamefont {Eremin}}]{lechermann2023electronic}%
  \BibitemOpen
  \bibfield  {author} {\bibinfo {author} {\bibfnamefont {F.}~\bibnamefont
  {Lechermann}}, \bibinfo {author} {\bibfnamefont {J.}~\bibnamefont {Gondolf}},
  \bibinfo {author} {\bibfnamefont {S.}~\bibnamefont {B\"otzel}},\ and\
  \bibinfo {author} {\bibfnamefont {I.~M.}\ \bibnamefont {Eremin}},\ }\bibfield
   {title} {\bibinfo {title} {{Electronic correlations and superconducting
  instability in ${\mathrm{La}}_{3}{\mathrm{Ni}}_{2}{\mathrm{O}}_{7}$ under
  high pressure}},\ }\href {https://doi.org/10.1103/PhysRevB.108.L201121}
  {\bibfield  {journal} {\bibinfo  {journal} {Phys. Rev. B}\ }\textbf {\bibinfo
  {volume} {108}},\ \bibinfo {pages} {L201121} (\bibinfo {year}
  {2023})}\BibitemShut {NoStop}%
\bibitem [{\citenamefont {Sakakibara}\ \emph
  {et~al.}(2024{\natexlab{a}})\citenamefont {Sakakibara}, \citenamefont
  {Kitamine}, \citenamefont {Ochi},\ and\ \citenamefont
  {Kuroki}}]{sakakibara2023possible}%
  \BibitemOpen
  \bibfield  {author} {\bibinfo {author} {\bibfnamefont {H.}~\bibnamefont
  {Sakakibara}}, \bibinfo {author} {\bibfnamefont {N.}~\bibnamefont
  {Kitamine}}, \bibinfo {author} {\bibfnamefont {M.}~\bibnamefont {Ochi}},\
  and\ \bibinfo {author} {\bibfnamefont {K.}~\bibnamefont {Kuroki}},\
  }\bibfield  {title} {\bibinfo {title} {{Possible High ${T}_{c}$
  Superconductivity in ${\mathrm{La}}_{3}{\mathrm{Ni}}_{2}{\mathrm{O}}_{7}$
  under High Pressure through Manifestation of a Nearly Half-Filled Bilayer
  Hubbard Model}},\ }\href {https://doi.org/10.1103/PhysRevLett.132.106002}
  {\bibfield  {journal} {\bibinfo  {journal} {Phys. Rev. Lett.}\ }\textbf
  {\bibinfo {volume} {132}},\ \bibinfo {pages} {106002} (\bibinfo {year}
  {2024}{\natexlab{a}})}\BibitemShut {NoStop}%
\bibitem [{\citenamefont {Gu}\ \emph {et~al.}(2023)\citenamefont {Gu},
  \citenamefont {Le}, \citenamefont {Yang}, \citenamefont {Wu},\ and\
  \citenamefont {Hu}}]{gu2023effective}%
  \BibitemOpen
  \bibfield  {author} {\bibinfo {author} {\bibfnamefont {Y.}~\bibnamefont
  {Gu}}, \bibinfo {author} {\bibfnamefont {C.}~\bibnamefont {Le}}, \bibinfo
  {author} {\bibfnamefont {Z.}~\bibnamefont {Yang}}, \bibinfo {author}
  {\bibfnamefont {X.}~\bibnamefont {Wu}},\ and\ \bibinfo {author}
  {\bibfnamefont {J.}~\bibnamefont {Hu}},\ }\href@noop {} {\bibinfo {title}
  {Effective model and pairing tendency in bilayer {Ni-based} superconductor
  {La$_3$Ni$_2$O$_7$}}} (\bibinfo {year} {2023}),\ \Eprint
  {https://arxiv.org/abs/2306.07275} {arXiv:2306.07275 [cond-mat.supr-con]}
  \BibitemShut {NoStop}%
\bibitem [{\citenamefont {Shen}\ \emph {et~al.}(2023)\citenamefont {Shen},
  \citenamefont {Qin},\ and\ \citenamefont {Zhang}}]{shen2023effective}%
  \BibitemOpen
  \bibfield  {author} {\bibinfo {author} {\bibfnamefont {Y.}~\bibnamefont
  {Shen}}, \bibinfo {author} {\bibfnamefont {M.}~\bibnamefont {Qin}},\ and\
  \bibinfo {author} {\bibfnamefont {G.-M.}\ \bibnamefont {Zhang}},\ }\bibfield
  {title} {\bibinfo {title} {Effective bi-layer model hamiltonian and
  density-matrix renormalization group study for the high-{$T_c$}
  superconductivity in {La$_3$Ni$_2$O$_7$} under high pressure},\ }\href
  {https://doi.org/10.1088/0256-307X/40/12/127401} {\bibfield  {journal}
  {\bibinfo  {journal} {Chinese Physics Letters}\ }\textbf {\bibinfo {volume}
  {40}},\ \bibinfo {pages} {127401} (\bibinfo {year} {2023})}\BibitemShut
  {NoStop}%
\bibitem [{\citenamefont {Christiansson}\ \emph {et~al.}(2023)\citenamefont
  {Christiansson}, \citenamefont {Petocchi},\ and\ \citenamefont
  {Werner}}]{christiansson2023correlated}%
  \BibitemOpen
  \bibfield  {author} {\bibinfo {author} {\bibfnamefont {V.}~\bibnamefont
  {Christiansson}}, \bibinfo {author} {\bibfnamefont {F.}~\bibnamefont
  {Petocchi}},\ and\ \bibinfo {author} {\bibfnamefont {P.}~\bibnamefont
  {Werner}},\ }\bibfield  {title} {\bibinfo {title} {{Correlated Electronic
  Structure of ${\mathrm{La}}_{3}{\text{Ni}}_{2}{\mathrm{O}}_{7}$ under
  Pressure}},\ }\href {https://doi.org/10.1103/PhysRevLett.131.206501}
  {\bibfield  {journal} {\bibinfo  {journal} {Phys. Rev. Lett.}\ }\textbf
  {\bibinfo {volume} {131}},\ \bibinfo {pages} {206501} (\bibinfo {year}
  {2023})}\BibitemShut {NoStop}%
\bibitem [{\citenamefont {Shilenko}\ and\ \citenamefont
  {Leonov}(2023)}]{Shilenko2023Correlated}%
  \BibitemOpen
  \bibfield  {author} {\bibinfo {author} {\bibfnamefont {D.~A.}\ \bibnamefont
  {Shilenko}}\ and\ \bibinfo {author} {\bibfnamefont {I.~V.}\ \bibnamefont
  {Leonov}},\ }\bibfield  {title} {\bibinfo {title} {Correlated electronic
  structure, orbital-selective behavior, and magnetic correlations in
  double-layer {${\mathrm{La}}_{3}{\mathrm{Ni}}_{2}{\mathrm{O}}_{7}$} under
  pressure},\ }\href {https://doi.org/10.1103/PhysRevB.108.125105} {\bibfield
  {journal} {\bibinfo  {journal} {Phys. Rev. B}\ }\textbf {\bibinfo {volume}
  {108}},\ \bibinfo {pages} {125105} (\bibinfo {year} {2023})}\BibitemShut
  {NoStop}%
\bibitem [{\citenamefont {W\'{u}}\ \emph {et~al.}(2024)\citenamefont {W\'{u}},
  \citenamefont {Luo}, \citenamefont {Yao},\ and\ \citenamefont
  {Wang}}]{wu2023charge}%
  \BibitemOpen
  \bibfield  {author} {\bibinfo {author} {\bibfnamefont {W.}~\bibnamefont
  {W\'{u}}}, \bibinfo {author} {\bibfnamefont {Z.}~\bibnamefont {Luo}},
  \bibinfo {author} {\bibfnamefont {D.-X.}\ \bibnamefont {Yao}},\ and\ \bibinfo
  {author} {\bibfnamefont {M.}~\bibnamefont {Wang}},\ }\bibfield  {title}
  {\bibinfo {title} {Superexchange and charge transfer in the nickelate
  superconductor {$\mathrm{La_3Ni_2O_7}$} under pressure},\ }\href
  {https://doi.org/https://doi.org/10.1007/s11433-023-2300-4} {\bibfield
  {journal} {\bibinfo  {journal} {SCIENCE CHINA Physics, Mechanics \&
  Astronomy}\ }\textbf {\bibinfo {volume} {67}},\ \bibinfo {pages} {117402}
  (\bibinfo {year} {2024})}\BibitemShut {NoStop}%
\bibitem [{\citenamefont {Cao}\ and\ \citenamefont {Yang}(2024)}]{cao2023flat}%
  \BibitemOpen
  \bibfield  {author} {\bibinfo {author} {\bibfnamefont {Y.}~\bibnamefont
  {Cao}}\ and\ \bibinfo {author} {\bibfnamefont {Y.-f.}\ \bibnamefont {Yang}},\
  }\bibfield  {title} {\bibinfo {title} {Flat bands promoted by hund's rule
  coupling in the candidate double-layer high-temperature superconductor
  {${\mathrm{La}}_{3}{\mathrm{Ni}}_{2}{\mathrm{O}}_{7}$} under high pressure},\
  }\href {https://doi.org/10.1103/PhysRevB.109.L081105} {\bibfield  {journal}
  {\bibinfo  {journal} {Phys. Rev. B}\ }\textbf {\bibinfo {volume} {109}},\
  \bibinfo {pages} {L081105} (\bibinfo {year} {2024})}\BibitemShut {NoStop}%
\bibitem [{\citenamefont {Chen}\ \emph {et~al.}(2023)\citenamefont {Chen},
  \citenamefont {Jiang}, \citenamefont {Li}, \citenamefont {Zhong},\ and\
  \citenamefont {Lu}}]{chen2023critical}%
  \BibitemOpen
  \bibfield  {author} {\bibinfo {author} {\bibfnamefont {X.}~\bibnamefont
  {Chen}}, \bibinfo {author} {\bibfnamefont {P.}~\bibnamefont {Jiang}},
  \bibinfo {author} {\bibfnamefont {J.}~\bibnamefont {Li}}, \bibinfo {author}
  {\bibfnamefont {Z.}~\bibnamefont {Zhong}},\ and\ \bibinfo {author}
  {\bibfnamefont {Y.}~\bibnamefont {Lu}},\ }\href@noop {} {\bibinfo {title}
  {Critical charge and spin instabilities in superconducting
  {La$_3$Ni$_2$O$_7$}}} (\bibinfo {year} {2023}),\ \Eprint
  {https://arxiv.org/abs/2307.07154} {arXiv:2307.07154 [cond-mat.supr-con]}
  \BibitemShut {NoStop}%
\bibitem [{\citenamefont {Liu}\ \emph {et~al.}(2023{\natexlab{b}})\citenamefont
  {Liu}, \citenamefont {Mei}, \citenamefont {Ye}, \citenamefont {Chen},\ and\
  \citenamefont {Yang}}]{liu2023spmwave}%
  \BibitemOpen
  \bibfield  {author} {\bibinfo {author} {\bibfnamefont {Y.-B.}\ \bibnamefont
  {Liu}}, \bibinfo {author} {\bibfnamefont {J.-W.}\ \bibnamefont {Mei}},
  \bibinfo {author} {\bibfnamefont {F.}~\bibnamefont {Ye}}, \bibinfo {author}
  {\bibfnamefont {W.-Q.}\ \bibnamefont {Chen}},\ and\ \bibinfo {author}
  {\bibfnamefont {F.}~\bibnamefont {Yang}},\ }\bibfield  {title} {\bibinfo
  {title} {{${\mathrm{s}}^{\ifmmode\pm\else\textpm\fi{}}$-Wave Pairing and the
  Destructive Role of Apical-Oxygen Deficiencies in
  ${\mathrm{La}}_{3}{\mathrm{Ni}}_{2}{\mathrm{O}}_{7}$ under Pressure}},\
  }\href {https://doi.org/10.1103/PhysRevLett.131.236002} {\bibfield  {journal}
  {\bibinfo  {journal} {Phys. Rev. Lett.}\ }\textbf {\bibinfo {volume} {131}},\
  \bibinfo {pages} {236002} (\bibinfo {year} {2023}{\natexlab{b}})}\BibitemShut
  {NoStop}%
\bibitem [{\citenamefont {Lu}\ \emph {et~al.}(2024)\citenamefont {Lu},
  \citenamefont {Pan}, \citenamefont {Yang},\ and\ \citenamefont
  {Wu}}]{lu2023interlayer}%
  \BibitemOpen
  \bibfield  {author} {\bibinfo {author} {\bibfnamefont {C.}~\bibnamefont
  {Lu}}, \bibinfo {author} {\bibfnamefont {Z.}~\bibnamefont {Pan}}, \bibinfo
  {author} {\bibfnamefont {F.}~\bibnamefont {Yang}},\ and\ \bibinfo {author}
  {\bibfnamefont {C.}~\bibnamefont {Wu}},\ }\bibfield  {title} {\bibinfo
  {title} {{Interlayer-Coupling-Driven High-Temperature Superconductivity in
  ${\mathrm{La}}_{3}{\mathrm{Ni}}_{2}{\mathrm{O}}_{7}$ under Pressure}},\
  }\href {https://doi.org/10.1103/PhysRevLett.132.146002} {\bibfield  {journal}
  {\bibinfo  {journal} {Phys. Rev. Lett.}\ }\textbf {\bibinfo {volume} {132}},\
  \bibinfo {pages} {146002} (\bibinfo {year} {2024})}\BibitemShut {NoStop}%
\bibitem [{\citenamefont {Zhang}\ \emph
  {et~al.}(2024{\natexlab{b}})\citenamefont {Zhang}, \citenamefont {Lin},
  \citenamefont {Moreo}, \citenamefont {Maier},\ and\ \citenamefont
  {Dagotto}}]{zhang2023structural}%
  \BibitemOpen
  \bibfield  {author} {\bibinfo {author} {\bibfnamefont {Y.}~\bibnamefont
  {Zhang}}, \bibinfo {author} {\bibfnamefont {L.-F.}\ \bibnamefont {Lin}},
  \bibinfo {author} {\bibfnamefont {A.}~\bibnamefont {Moreo}}, \bibinfo
  {author} {\bibfnamefont {T.~A.}\ \bibnamefont {Maier}},\ and\ \bibinfo
  {author} {\bibfnamefont {E.}~\bibnamefont {Dagotto}},\ }\bibfield  {title}
  {\bibinfo {title} {Structural phase transition, s±-wave pairing, and
  magnetic stripe order in bilayered superconductor la3ni2o7 under pressure},\
  }\href {https://doi.org/10.1038/s41467-024-46622-z} {\bibfield  {journal}
  {\bibinfo  {journal} {Nature Communications}\ }\textbf {\bibinfo {volume}
  {15}},\ \bibinfo {pages} {2470} (\bibinfo {year}
  {2024}{\natexlab{b}})}\BibitemShut {NoStop}%
\bibitem [{\citenamefont {Oh}\ and\ \citenamefont {Zhang}(2023)}]{oh2023type}%
  \BibitemOpen
  \bibfield  {author} {\bibinfo {author} {\bibfnamefont {H.}~\bibnamefont
  {Oh}}\ and\ \bibinfo {author} {\bibfnamefont {Y.-H.}\ \bibnamefont {Zhang}},\
  }\bibfield  {title} {\bibinfo {title} {{Type-II $t\ensuremath{-}J$ model and
  shared superexchange coupling from Hund's rule in superconducting
  ${\mathrm{La}}_{3}{\mathrm{Ni}}_{2}{\mathrm{O}}_{7}$}},\ }\href
  {https://doi.org/10.1103/PhysRevB.108.174511} {\bibfield  {journal} {\bibinfo
   {journal} {Phys. Rev. B}\ }\textbf {\bibinfo {volume} {108}},\ \bibinfo
  {pages} {174511} (\bibinfo {year} {2023})}\BibitemShut {NoStop}%
\bibitem [{\citenamefont {Liao}\ \emph {et~al.}(2023)\citenamefont {Liao},
  \citenamefont {Chen}, \citenamefont {Duan}, \citenamefont {Wang},
  \citenamefont {Liu}, \citenamefont {Yu},\ and\ \citenamefont
  {Si}}]{liao2023electron}%
  \BibitemOpen
  \bibfield  {author} {\bibinfo {author} {\bibfnamefont {Z.}~\bibnamefont
  {Liao}}, \bibinfo {author} {\bibfnamefont {L.}~\bibnamefont {Chen}}, \bibinfo
  {author} {\bibfnamefont {G.}~\bibnamefont {Duan}}, \bibinfo {author}
  {\bibfnamefont {Y.}~\bibnamefont {Wang}}, \bibinfo {author} {\bibfnamefont
  {C.}~\bibnamefont {Liu}}, \bibinfo {author} {\bibfnamefont {R.}~\bibnamefont
  {Yu}},\ and\ \bibinfo {author} {\bibfnamefont {Q.}~\bibnamefont {Si}},\
  }\bibfield  {title} {\bibinfo {title} {{Electron correlations and
  superconductivity in ${\mathrm{La}}_{3}{\mathrm{Ni}}_{2}{\mathrm{O}}_{7}$
  under pressure tuning}},\ }\href
  {https://doi.org/10.1103/PhysRevB.108.214522} {\bibfield  {journal} {\bibinfo
   {journal} {Phys. Rev. B}\ }\textbf {\bibinfo {volume} {108}},\ \bibinfo
  {pages} {214522} (\bibinfo {year} {2023})}\BibitemShut {NoStop}%
\bibitem [{\citenamefont {Qu}\ \emph {et~al.}(2024)\citenamefont {Qu},
  \citenamefont {Qu}, \citenamefont {Chen}, \citenamefont {Wu}, \citenamefont
  {Yang}, \citenamefont {Li},\ and\ \citenamefont {Su}}]{qu2023bilayer}%
  \BibitemOpen
  \bibfield  {author} {\bibinfo {author} {\bibfnamefont {X.-Z.}\ \bibnamefont
  {Qu}}, \bibinfo {author} {\bibfnamefont {D.-W.}\ \bibnamefont {Qu}}, \bibinfo
  {author} {\bibfnamefont {J.}~\bibnamefont {Chen}}, \bibinfo {author}
  {\bibfnamefont {C.}~\bibnamefont {Wu}}, \bibinfo {author} {\bibfnamefont
  {F.}~\bibnamefont {Yang}}, \bibinfo {author} {\bibfnamefont {W.}~\bibnamefont
  {Li}},\ and\ \bibinfo {author} {\bibfnamefont {G.}~\bibnamefont {Su}},\
  }\bibfield  {title} {\bibinfo {title} {Bilayer
  ${t\text{\ensuremath{-}}J\text{\ensuremath{-}}J}_{\ensuremath{\perp}}$ model
  and magnetically mediated pairing in the pressurized nickelate
  ${\mathrm{la}}_{3}{\mathrm{ni}}_{2}{\mathrm{o}}_{7}$},\ }\href
  {https://doi.org/10.1103/PhysRevLett.132.036502} {\bibfield  {journal}
  {\bibinfo  {journal} {Phys. Rev. Lett.}\ }\textbf {\bibinfo {volume} {132}},\
  \bibinfo {pages} {036502} (\bibinfo {year} {2024})}\BibitemShut {NoStop}%
\bibitem [{\citenamefont {Yang}\ \emph
  {et~al.}(2023{\natexlab{c}})\citenamefont {Yang}, \citenamefont {Zhang},\
  and\ \citenamefont {Zhang}}]{yang2023minimal}%
  \BibitemOpen
  \bibfield  {author} {\bibinfo {author} {\bibfnamefont {Y.-f.}\ \bibnamefont
  {Yang}}, \bibinfo {author} {\bibfnamefont {G.-M.}\ \bibnamefont {Zhang}},\
  and\ \bibinfo {author} {\bibfnamefont {F.-C.}\ \bibnamefont {Zhang}},\
  }\bibfield  {title} {\bibinfo {title} {{Interlayer valence bonds and
  two-component theory for high-${T}_{c}$ superconductivity of
  ${\mathrm{La}}_{3}{\mathrm{Ni}}_{2}{\mathrm{O}}_{7}$ under pressure}},\
  }\href {https://doi.org/10.1103/PhysRevB.108.L201108} {\bibfield  {journal}
  {\bibinfo  {journal} {Phys. Rev. B}\ }\textbf {\bibinfo {volume} {108}},\
  \bibinfo {pages} {L201108} (\bibinfo {year}
  {2023}{\natexlab{c}})}\BibitemShut {NoStop}%
\bibitem [{\citenamefont {Jiang}\ \emph
  {et~al.}(2024{\natexlab{a}})\citenamefont {Jiang}, \citenamefont {Wang},\
  and\ \citenamefont {Zhang}}]{jiang2023high}%
  \BibitemOpen
  \bibfield  {author} {\bibinfo {author} {\bibfnamefont {K.}~\bibnamefont
  {Jiang}}, \bibinfo {author} {\bibfnamefont {Z.}~\bibnamefont {Wang}},\ and\
  \bibinfo {author} {\bibfnamefont {F.-C.}\ \bibnamefont {Zhang}},\ }\bibfield
  {title} {\bibinfo {title} {High-temperature superconductivity in
  {La$_3$Ni$_2$O$_7$}},\ }\href {https://doi.org/10.1088/0256-307X/41/1/017402}
  {\bibfield  {journal} {\bibinfo  {journal} {Chinese Physics Letters}\
  }\textbf {\bibinfo {volume} {41}},\ \bibinfo {pages} {017402} (\bibinfo
  {year} {2024}{\natexlab{a}})}\BibitemShut {NoStop}%
\bibitem [{\citenamefont {Zhang}\ \emph
  {et~al.}(2023{\natexlab{c}})\citenamefont {Zhang}, \citenamefont {Lin},
  \citenamefont {Moreo}, \citenamefont {Maier},\ and\ \citenamefont
  {Dagotto}}]{zhang2023trends}%
  \BibitemOpen
  \bibfield  {author} {\bibinfo {author} {\bibfnamefont {Y.}~\bibnamefont
  {Zhang}}, \bibinfo {author} {\bibfnamefont {L.-F.}\ \bibnamefont {Lin}},
  \bibinfo {author} {\bibfnamefont {A.}~\bibnamefont {Moreo}}, \bibinfo
  {author} {\bibfnamefont {T.~A.}\ \bibnamefont {Maier}},\ and\ \bibinfo
  {author} {\bibfnamefont {E.}~\bibnamefont {Dagotto}},\ }\bibfield  {title}
  {\bibinfo {title} {Trends in electronic structures and
  ${s}_{\ifmmode\pm\else\textpm\fi{}}$-wave pairing for the rare-earth series
  in bilayer nickelate superconductor
  {${R}_{3}{\mathrm{Ni}}_{2}{\mathrm{O}}_{7}$}},\ }\href
  {https://doi.org/10.1103/PhysRevB.108.165141} {\bibfield  {journal} {\bibinfo
   {journal} {Phys. Rev. B}\ }\textbf {\bibinfo {volume} {108}},\ \bibinfo
  {pages} {165141} (\bibinfo {year} {2023}{\natexlab{c}})}\BibitemShut
  {NoStop}%
\bibitem [{\citenamefont {Huang}\ \emph {et~al.}(2023)\citenamefont {Huang},
  \citenamefont {Wang},\ and\ \citenamefont {Zhou}}]{huang2023impurity}%
  \BibitemOpen
  \bibfield  {author} {\bibinfo {author} {\bibfnamefont {J.}~\bibnamefont
  {Huang}}, \bibinfo {author} {\bibfnamefont {Z.~D.}\ \bibnamefont {Wang}},\
  and\ \bibinfo {author} {\bibfnamefont {T.}~\bibnamefont {Zhou}},\ }\bibfield
  {title} {\bibinfo {title} {Impurity and vortex states in the bilayer
  high-temperature superconductor
  {${\mathrm{La}}_{3}{\mathrm{Ni}}_{2}{\mathrm{O}}_{7}$}},\ }\href
  {https://doi.org/10.1103/PhysRevB.108.174501} {\bibfield  {journal} {\bibinfo
   {journal} {Phys. Rev. B}\ }\textbf {\bibinfo {volume} {108}},\ \bibinfo
  {pages} {174501} (\bibinfo {year} {2023})}\BibitemShut {NoStop}%
\bibitem [{\citenamefont {Qin}\ and\ \citenamefont
  {Yang}(2023)}]{qin2023hightc}%
  \BibitemOpen
  \bibfield  {author} {\bibinfo {author} {\bibfnamefont {Q.}~\bibnamefont
  {Qin}}\ and\ \bibinfo {author} {\bibfnamefont {Y.-F.}\ \bibnamefont {Yang}},\
  }\bibfield  {title} {\bibinfo {title} {High-${T}_{c}$ superconductivity by
  mobilizing local spin singlets and possible route to higher ${T}_{c}$ in
  pressurized {${\mathrm{La}}_{3}{\mathrm{Ni}}_{2}{\mathrm{O}}_{7}$}},\ }\href
  {https://doi.org/10.1103/PhysRevB.108.L140504} {\bibfield  {journal}
  {\bibinfo  {journal} {Phys. Rev. B}\ }\textbf {\bibinfo {volume} {108}},\
  \bibinfo {pages} {L140504} (\bibinfo {year} {2023})}\BibitemShut {NoStop}%
\bibitem [{\citenamefont {Tian}\ \emph {et~al.}(2023)\citenamefont {Tian},
  \citenamefont {Chen}, \citenamefont {Wang}, \citenamefont {He},\ and\
  \citenamefont {Lu}}]{tian2023correlation}%
  \BibitemOpen
  \bibfield  {author} {\bibinfo {author} {\bibfnamefont {Y.-H.}\ \bibnamefont
  {Tian}}, \bibinfo {author} {\bibfnamefont {Y.}~\bibnamefont {Chen}}, \bibinfo
  {author} {\bibfnamefont {J.-M.}\ \bibnamefont {Wang}}, \bibinfo {author}
  {\bibfnamefont {R.-Q.}\ \bibnamefont {He}},\ and\ \bibinfo {author}
  {\bibfnamefont {Z.-Y.}\ \bibnamefont {Lu}},\ }\href@noop {} {\bibinfo {title}
  {Correlation effects and concomitant two-orbital $s_\pm$-wave
  superconductivity in {La$_3$Ni$_2$O$_7$} under high pressure}} (\bibinfo
  {year} {2023}),\ \Eprint {https://arxiv.org/abs/2308.09698} {arXiv:2308.09698
  [cond-mat.supr-con]} \BibitemShut {NoStop}%
\bibitem [{\citenamefont {Lu}\ \emph {et~al.}(2023{\natexlab{a}})\citenamefont
  {Lu}, \citenamefont {Li}, \citenamefont {Zeng}, \citenamefont {Hou},
  \citenamefont {Wang}, \citenamefont {Yang},\ and\ \citenamefont
  {You}}]{lu2023superconductivity}%
  \BibitemOpen
  \bibfield  {author} {\bibinfo {author} {\bibfnamefont {D.-C.}\ \bibnamefont
  {Lu}}, \bibinfo {author} {\bibfnamefont {M.}~\bibnamefont {Li}}, \bibinfo
  {author} {\bibfnamefont {Z.-Y.}\ \bibnamefont {Zeng}}, \bibinfo {author}
  {\bibfnamefont {W.}~\bibnamefont {Hou}}, \bibinfo {author} {\bibfnamefont
  {J.}~\bibnamefont {Wang}}, \bibinfo {author} {\bibfnamefont {F.}~\bibnamefont
  {Yang}},\ and\ \bibinfo {author} {\bibfnamefont {Y.-Z.}\ \bibnamefont
  {You}},\ }\href@noop {} {\bibinfo {title} {Superconductivity from doping
  symmetric mass generation insulators: Application to {La$_3$Ni$_2$O$_7$}
  under pressure}} (\bibinfo {year} {2023}{\natexlab{a}}),\ \Eprint
  {https://arxiv.org/abs/2308.11195} {arXiv:2308.11195 [cond-mat.str-el]}
  \BibitemShut {NoStop}%
\bibitem [{\citenamefont {Jiang}\ \emph
  {et~al.}(2024{\natexlab{b}})\citenamefont {Jiang}, \citenamefont {Hou},
  \citenamefont {Fan}, \citenamefont {Lang},\ and\ \citenamefont
  {Ku}}]{jiang2023pressure}%
  \BibitemOpen
  \bibfield  {author} {\bibinfo {author} {\bibfnamefont {R.}~\bibnamefont
  {Jiang}}, \bibinfo {author} {\bibfnamefont {J.}~\bibnamefont {Hou}}, \bibinfo
  {author} {\bibfnamefont {Z.}~\bibnamefont {Fan}}, \bibinfo {author}
  {\bibfnamefont {Z.-J.}\ \bibnamefont {Lang}},\ and\ \bibinfo {author}
  {\bibfnamefont {W.}~\bibnamefont {Ku}},\ }\bibfield  {title} {\bibinfo
  {title} {Pressure driven fractionalization of ionic spins results in
  cupratelike high-${T}_{c}$ superconductivity in
  {${\mathrm{La}}_{3}{\mathrm{Ni}}_{2}{\mathrm{O}}_{7}$}},\ }\href
  {https://doi.org/10.1103/PhysRevLett.132.126503} {\bibfield  {journal}
  {\bibinfo  {journal} {Phys. Rev. Lett.}\ }\textbf {\bibinfo {volume} {132}},\
  \bibinfo {pages} {126503} (\bibinfo {year} {2024}{\natexlab{b}})}\BibitemShut
  {NoStop}%
\bibitem [{\citenamefont {Kitamine}\ \emph {et~al.}(2023)\citenamefont
  {Kitamine}, \citenamefont {Ochi},\ and\ \citenamefont
  {Kuroki}}]{kitamine2023theoretical}%
  \BibitemOpen
  \bibfield  {author} {\bibinfo {author} {\bibfnamefont {N.}~\bibnamefont
  {Kitamine}}, \bibinfo {author} {\bibfnamefont {M.}~\bibnamefont {Ochi}},\
  and\ \bibinfo {author} {\bibfnamefont {K.}~\bibnamefont {Kuroki}},\
  }\href@noop {} {\bibinfo {title} {Theoretical designing of multiband
  nickelate and palladate superconductors with $d^{8+\delta}$ configuration}}
  (\bibinfo {year} {2023}),\ \Eprint {https://arxiv.org/abs/2308.12750}
  {arXiv:2308.12750 [cond-mat.supr-con]} \BibitemShut {NoStop}%
\bibitem [{\citenamefont {Luo}\ \emph {et~al.}(2023{\natexlab{b}})\citenamefont
  {Luo}, \citenamefont {Lv}, \citenamefont {Wang}, \citenamefont {Wú},\ and\
  \citenamefont {Yao}}]{luo2023hightc}%
  \BibitemOpen
  \bibfield  {author} {\bibinfo {author} {\bibfnamefont {Z.}~\bibnamefont
  {Luo}}, \bibinfo {author} {\bibfnamefont {B.}~\bibnamefont {Lv}}, \bibinfo
  {author} {\bibfnamefont {M.}~\bibnamefont {Wang}}, \bibinfo {author}
  {\bibfnamefont {W.}~\bibnamefont {Wú}},\ and\ \bibinfo {author}
  {\bibfnamefont {D.-X.}\ \bibnamefont {Yao}},\ }\href@noop {} {\bibinfo
  {title} {{High-T$_C$} superconductivity in {$\mathrm{La_3Ni_2O_7}$} based on
  the bilayer two-orbital {$t$-$J$} model}} (\bibinfo {year}
  {2023}{\natexlab{b}}),\ \Eprint {https://arxiv.org/abs/2308.16564}
  {arXiv:2308.16564 [cond-mat.supr-con]} \BibitemShut {NoStop}%
\bibitem [{\citenamefont {Zhang}\ \emph
  {et~al.}(2023{\natexlab{d}})\citenamefont {Zhang}, \citenamefont {Zhang},
  \citenamefont {You},\ and\ \citenamefont {Weng}}]{zhang2023strong}%
  \BibitemOpen
  \bibfield  {author} {\bibinfo {author} {\bibfnamefont {J.-X.}\ \bibnamefont
  {Zhang}}, \bibinfo {author} {\bibfnamefont {H.-K.}\ \bibnamefont {Zhang}},
  \bibinfo {author} {\bibfnamefont {Y.-Z.}\ \bibnamefont {You}},\ and\ \bibinfo
  {author} {\bibfnamefont {Z.-Y.}\ \bibnamefont {Weng}},\ }\href@noop {}
  {\bibinfo {title} {Strong pairing originated from an emergent
  {$\mathbb{Z}_2$} berry phase in $\mathbf{La}_3 \mathbf{Ni}_2 \mathbf{O}_7$}}
  (\bibinfo {year} {2023}{\natexlab{d}}),\ \Eprint
  {https://arxiv.org/abs/2309.05726} {arXiv:2309.05726 [cond-mat.str-el]}
  \BibitemShut {NoStop}%
\bibitem [{\citenamefont {Pan}\ \emph {et~al.}(2023)\citenamefont {Pan},
  \citenamefont {Lu}, \citenamefont {Yang},\ and\ \citenamefont
  {Wu}}]{pan2023effect}%
  \BibitemOpen
  \bibfield  {author} {\bibinfo {author} {\bibfnamefont {Z.}~\bibnamefont
  {Pan}}, \bibinfo {author} {\bibfnamefont {C.}~\bibnamefont {Lu}}, \bibinfo
  {author} {\bibfnamefont {F.}~\bibnamefont {Yang}},\ and\ \bibinfo {author}
  {\bibfnamefont {C.}~\bibnamefont {Wu}},\ }\href@noop {} {\bibinfo {title}
  {Effect of rare-earth element substitution in superconducting
  {R$_3$Ni$_2$O$_7$} under pressure}} (\bibinfo {year} {2023}),\ \Eprint
  {https://arxiv.org/abs/2309.06173} {arXiv:2309.06173 [cond-mat.supr-con]}
  \BibitemShut {NoStop}%
\bibitem [{\citenamefont {Sakakibara}\ \emph
  {et~al.}(2024{\natexlab{b}})\citenamefont {Sakakibara}, \citenamefont {Ochi},
  \citenamefont {Nagata}, \citenamefont {Ueki}, \citenamefont {Sakurai},
  \citenamefont {Matsumoto}, \citenamefont {Terashima}, \citenamefont {Hirose},
  \citenamefont {Ohta}, \citenamefont {Kato}, \citenamefont {Takano},\ and\
  \citenamefont {Kuroki}}]{sakakibara2023theoretical}%
  \BibitemOpen
  \bibfield  {author} {\bibinfo {author} {\bibfnamefont {H.}~\bibnamefont
  {Sakakibara}}, \bibinfo {author} {\bibfnamefont {M.}~\bibnamefont {Ochi}},
  \bibinfo {author} {\bibfnamefont {H.}~\bibnamefont {Nagata}}, \bibinfo
  {author} {\bibfnamefont {Y.}~\bibnamefont {Ueki}}, \bibinfo {author}
  {\bibfnamefont {H.}~\bibnamefont {Sakurai}}, \bibinfo {author} {\bibfnamefont
  {R.}~\bibnamefont {Matsumoto}}, \bibinfo {author} {\bibfnamefont
  {K.}~\bibnamefont {Terashima}}, \bibinfo {author} {\bibfnamefont
  {K.}~\bibnamefont {Hirose}}, \bibinfo {author} {\bibfnamefont
  {H.}~\bibnamefont {Ohta}}, \bibinfo {author} {\bibfnamefont {M.}~\bibnamefont
  {Kato}}, \bibinfo {author} {\bibfnamefont {Y.}~\bibnamefont {Takano}},\ and\
  \bibinfo {author} {\bibfnamefont {K.}~\bibnamefont {Kuroki}},\ }\bibfield
  {title} {\bibinfo {title} {{Theoretical analysis on the possibility of
  superconductivity in the trilayer Ruddlesden-Popper nickelate
  ${\mathrm{La}}_{4}{\mathrm{Ni}}_{3}{\mathrm{O}}_{10}$ under pressure and its
  experimental examination: Comparison with
  ${\mathrm{La}}_{3}{\mathrm{Ni}}_{2}{\mathrm{O}}_{7}$}},\ }\href
  {https://doi.org/10.1103/PhysRevB.109.144511} {\bibfield  {journal} {\bibinfo
   {journal} {Phys. Rev. B}\ }\textbf {\bibinfo {volume} {109}},\ \bibinfo
  {pages} {144511} (\bibinfo {year} {2024}{\natexlab{b}})}\BibitemShut
  {NoStop}%
\bibitem [{\citenamefont {Lange}\ \emph
  {et~al.}(2023{\natexlab{a}})\citenamefont {Lange}, \citenamefont {Homeier},
  \citenamefont {Demler}, \citenamefont {Schollwöck}, \citenamefont {Bohrdt},\
  and\ \citenamefont {Grusdt}}]{lange2023pairing}%
  \BibitemOpen
  \bibfield  {author} {\bibinfo {author} {\bibfnamefont {H.}~\bibnamefont
  {Lange}}, \bibinfo {author} {\bibfnamefont {L.}~\bibnamefont {Homeier}},
  \bibinfo {author} {\bibfnamefont {E.}~\bibnamefont {Demler}}, \bibinfo
  {author} {\bibfnamefont {U.}~\bibnamefont {Schollwöck}}, \bibinfo {author}
  {\bibfnamefont {A.}~\bibnamefont {Bohrdt}},\ and\ \bibinfo {author}
  {\bibfnamefont {F.}~\bibnamefont {Grusdt}},\ }\href@noop {} {\bibinfo {title}
  {Pairing dome from an emergent {Feshbach} resonance in a strongly repulsive
  bilayer model}} (\bibinfo {year} {2023}{\natexlab{a}}),\ \Eprint
  {https://arxiv.org/abs/2309.13040} {arXiv:2309.13040 [cond-mat.str-el]}
  \BibitemShut {NoStop}%
\bibitem [{\citenamefont {Geisler}\ \emph {et~al.}(2023)\citenamefont
  {Geisler}, \citenamefont {Hamlin}, \citenamefont {Stewart}, \citenamefont
  {Hennig},\ and\ \citenamefont {Hirschfeld}}]{geisler2023structural}%
  \BibitemOpen
  \bibfield  {author} {\bibinfo {author} {\bibfnamefont {B.}~\bibnamefont
  {Geisler}}, \bibinfo {author} {\bibfnamefont {J.~J.}\ \bibnamefont {Hamlin}},
  \bibinfo {author} {\bibfnamefont {G.~R.}\ \bibnamefont {Stewart}}, \bibinfo
  {author} {\bibfnamefont {R.~G.}\ \bibnamefont {Hennig}},\ and\ \bibinfo
  {author} {\bibfnamefont {P.~J.}\ \bibnamefont {Hirschfeld}},\ }\href@noop {}
  {\bibinfo {title} {Structural transitions, octahedral rotations, and
  electronic properties of {$A_3$Ni$_2$O$_7$} rare-earth nickelates under high
  pressure}} (\bibinfo {year} {2023}),\ \Eprint
  {https://arxiv.org/abs/2309.15078} {arXiv:2309.15078 [cond-mat.supr-con]}
  \BibitemShut {NoStop}%
\bibitem [{\citenamefont {Yang}\ \emph
  {et~al.}(2023{\natexlab{d}})\citenamefont {Yang}, \citenamefont {Oh},\ and\
  \citenamefont {Zhang}}]{yang2023strong}%
  \BibitemOpen
  \bibfield  {author} {\bibinfo {author} {\bibfnamefont {H.}~\bibnamefont
  {Yang}}, \bibinfo {author} {\bibfnamefont {H.}~\bibnamefont {Oh}},\ and\
  \bibinfo {author} {\bibfnamefont {Y.-H.}\ \bibnamefont {Zhang}},\ }\href@noop
  {} {\bibinfo {title} {Strong pairing from doping-induced {Feshbach} resonance
  and second {Fermi} liquid through doping a bilayer spin-one {Mott} insulator:
  application to {La$_3$Ni$_2$O$_7$}}} (\bibinfo {year} {2023}{\natexlab{d}}),\
  \Eprint {https://arxiv.org/abs/2309.15095} {arXiv:2309.15095
  [cond-mat.str-el]} \BibitemShut {NoStop}%
\bibitem [{\citenamefont {Rhodes}\ and\ \citenamefont
  {Wahl}(2024)}]{rhodes2023structural}%
  \BibitemOpen
  \bibfield  {author} {\bibinfo {author} {\bibfnamefont {L.~C.}\ \bibnamefont
  {Rhodes}}\ and\ \bibinfo {author} {\bibfnamefont {P.}~\bibnamefont {Wahl}},\
  }\bibfield  {title} {\bibinfo {title} {{Structural routes to stabilize
  superconducting ${\mathrm{La}}_{3}{\mathrm{Ni}}_{2}{\mathrm{O}}_{7}$ at
  ambient pressure}},\ }\href
  {https://doi.org/10.1103/PhysRevMaterials.8.044801} {\bibfield  {journal}
  {\bibinfo  {journal} {Phys. Rev. Mater.}\ }\textbf {\bibinfo {volume} {8}},\
  \bibinfo {pages} {044801} (\bibinfo {year} {2024})}\BibitemShut {NoStop}%
\bibitem [{\citenamefont {Lange}\ \emph
  {et~al.}(2023{\natexlab{b}})\citenamefont {Lange}, \citenamefont {Homeier},
  \citenamefont {Demler}, \citenamefont {Schollwöck}, \citenamefont {Grusdt},\
  and\ \citenamefont {Bohrdt}}]{lange2023feshbach}%
  \BibitemOpen
  \bibfield  {author} {\bibinfo {author} {\bibfnamefont {H.}~\bibnamefont
  {Lange}}, \bibinfo {author} {\bibfnamefont {L.}~\bibnamefont {Homeier}},
  \bibinfo {author} {\bibfnamefont {E.}~\bibnamefont {Demler}}, \bibinfo
  {author} {\bibfnamefont {U.}~\bibnamefont {Schollwöck}}, \bibinfo {author}
  {\bibfnamefont {F.}~\bibnamefont {Grusdt}},\ and\ \bibinfo {author}
  {\bibfnamefont {A.}~\bibnamefont {Bohrdt}},\ }\href@noop {} {\bibinfo {title}
  {{Feshbach} resonance in a strongly repulsive bilayer model: a possible
  scenario for bilayer nickelate superconductors}} (\bibinfo {year}
  {2023}{\natexlab{b}}),\ \Eprint {https://arxiv.org/abs/2309.15843}
  {arXiv:2309.15843 [cond-mat.str-el]} \BibitemShut {NoStop}%
\bibitem [{\citenamefont {LaBollita}\ \emph {et~al.}(2023)\citenamefont
  {LaBollita}, \citenamefont {Pardo}, \citenamefont {Norman},\ and\
  \citenamefont {Botana}}]{labollita2023electronic}%
  \BibitemOpen
  \bibfield  {author} {\bibinfo {author} {\bibfnamefont {H.}~\bibnamefont
  {LaBollita}}, \bibinfo {author} {\bibfnamefont {V.}~\bibnamefont {Pardo}},
  \bibinfo {author} {\bibfnamefont {M.~R.}\ \bibnamefont {Norman}},\ and\
  \bibinfo {author} {\bibfnamefont {A.~S.}\ \bibnamefont {Botana}},\
  }\href@noop {} {\bibinfo {title} {Electronic structure and magnetic
  properties of {La$_{3}$Ni$_{2}$O$_{7}$} under pressure}} (\bibinfo {year}
  {2023}),\ \Eprint {https://arxiv.org/abs/2309.17279} {arXiv:2309.17279
  [cond-mat.str-el]} \BibitemShut {NoStop}%
\bibitem [{\citenamefont {Kumar}\ \emph {et~al.}(2023)\citenamefont {Kumar},
  \citenamefont {Melnick},\ and\ \citenamefont {Kotliar}}]{kumar2023softening}%
  \BibitemOpen
  \bibfield  {author} {\bibinfo {author} {\bibfnamefont {U.}~\bibnamefont
  {Kumar}}, \bibinfo {author} {\bibfnamefont {C.}~\bibnamefont {Melnick}},\
  and\ \bibinfo {author} {\bibfnamefont {G.}~\bibnamefont {Kotliar}},\
  }\href@noop {} {\bibinfo {title} {Softening of $dd$ excitation in the
  resonant inelastic x-ray scattering spectra as a signature of {Hund's}
  coupling in nickelates}} (\bibinfo {year} {2023}),\ \Eprint
  {https://arxiv.org/abs/2310.00983} {arXiv:2310.00983 [cond-mat.str-el]}
  \BibitemShut {NoStop}%
\bibitem [{\citenamefont {Kaneko}\ \emph {et~al.}(2024)\citenamefont {Kaneko},
  \citenamefont {Sakakibara}, \citenamefont {Ochi},\ and\ \citenamefont
  {Kuroki}}]{kaneko2023pair}%
  \BibitemOpen
  \bibfield  {author} {\bibinfo {author} {\bibfnamefont {T.}~\bibnamefont
  {Kaneko}}, \bibinfo {author} {\bibfnamefont {H.}~\bibnamefont {Sakakibara}},
  \bibinfo {author} {\bibfnamefont {M.}~\bibnamefont {Ochi}},\ and\ \bibinfo
  {author} {\bibfnamefont {K.}~\bibnamefont {Kuroki}},\ }\bibfield  {title}
  {\bibinfo {title} {Pair correlations in the two-orbital hubbard ladder:
  Implications for superconductivity in the bilayer nickelate
  ${\mathrm{la}}_{3}{\mathrm{ni}}_{2}{\mathrm{o}}_{7}$},\ }\href
  {https://doi.org/10.1103/PhysRevB.109.045154} {\bibfield  {journal} {\bibinfo
   {journal} {Phys. Rev. B}\ }\textbf {\bibinfo {volume} {109}},\ \bibinfo
  {pages} {045154} (\bibinfo {year} {2024})}\BibitemShut {NoStop}%
\bibitem [{\citenamefont {Lu}\ \emph {et~al.}(2023{\natexlab{b}})\citenamefont
  {Lu}, \citenamefont {Pan}, \citenamefont {Yang},\ and\ \citenamefont
  {Wu}}]{lu2023interplay}%
  \BibitemOpen
  \bibfield  {author} {\bibinfo {author} {\bibfnamefont {C.}~\bibnamefont
  {Lu}}, \bibinfo {author} {\bibfnamefont {Z.}~\bibnamefont {Pan}}, \bibinfo
  {author} {\bibfnamefont {F.}~\bibnamefont {Yang}},\ and\ \bibinfo {author}
  {\bibfnamefont {C.}~\bibnamefont {Wu}},\ }\href@noop {} {\bibinfo {title}
  {Interplay of two {$E_g$} orbitals in superconducting {La$_3$Ni$_2$O$_7$}
  under pressure}} (\bibinfo {year} {2023}{\natexlab{b}}),\ \Eprint
  {https://arxiv.org/abs/2310.02915} {arXiv:2310.02915 [cond-mat.supr-con]}
  \BibitemShut {NoStop}%
\bibitem [{\citenamefont {Ryee}\ \emph {et~al.}(2023)\citenamefont {Ryee},
  \citenamefont {Witt},\ and\ \citenamefont {Wehling}}]{ryee2023critical}%
  \BibitemOpen
  \bibfield  {author} {\bibinfo {author} {\bibfnamefont {S.}~\bibnamefont
  {Ryee}}, \bibinfo {author} {\bibfnamefont {N.}~\bibnamefont {Witt}},\ and\
  \bibinfo {author} {\bibfnamefont {T.~O.}\ \bibnamefont {Wehling}},\
  }\href@noop {} {\bibinfo {title} {Critical role of interlayer dimer
  correlations in the superconductivity of {La$_3$Ni$_2$O$_7$}}} (\bibinfo
  {year} {2023}),\ \Eprint {https://arxiv.org/abs/2310.17465} {arXiv:2310.17465
  [cond-mat.supr-con]} \BibitemShut {NoStop}%
\bibitem [{\citenamefont {Schlömer}\ \emph {et~al.}(2023)\citenamefont
  {Schlömer}, \citenamefont {Schollwöck}, \citenamefont {Grusdt},\ and\
  \citenamefont {Bohrdt}}]{schlomer2023superconductivity}%
  \BibitemOpen
  \bibfield  {author} {\bibinfo {author} {\bibfnamefont {H.}~\bibnamefont
  {Schlömer}}, \bibinfo {author} {\bibfnamefont {U.}~\bibnamefont
  {Schollwöck}}, \bibinfo {author} {\bibfnamefont {F.}~\bibnamefont
  {Grusdt}},\ and\ \bibinfo {author} {\bibfnamefont {A.}~\bibnamefont
  {Bohrdt}},\ }\href@noop {} {\bibinfo {title} {Superconductivity in the
  pressurized nickelate {La$_3$Ni$_2$O$_7$} in the vicinity of a {BEC-BCS}
  crossover}} (\bibinfo {year} {2023}),\ \Eprint
  {https://arxiv.org/abs/2311.03349} {arXiv:2311.03349 [cond-mat.str-el]}
  \BibitemShut {NoStop}%
\bibitem [{\citenamefont {Verstraete}\ and\ \citenamefont
  {Cirac}(2004)}]{Verstraete2004renorm}%
  \BibitemOpen
  \bibfield  {author} {\bibinfo {author} {\bibfnamefont {F.}~\bibnamefont
  {Verstraete}}\ and\ \bibinfo {author} {\bibfnamefont {J.~I.}\ \bibnamefont
  {Cirac}},\ }\href@noop {} {\bibinfo {title} {Renormalization algorithms for
  quantum-many body systems in two and higher dimensions}} (\bibinfo {year}
  {2004}),\ \Eprint {https://arxiv.org/abs/cond-mat/0407066}
  {arXiv:cond-mat/0407066 [cond-mat.str-el]} \BibitemShut {NoStop}%
\bibitem [{\citenamefont {Jordan}\ \emph {et~al.}(2008)\citenamefont {Jordan},
  \citenamefont {Or\'us}, \citenamefont {Vidal}, \citenamefont {Verstraete},\
  and\ \citenamefont {Cirac}}]{Jordan2008Classical}%
  \BibitemOpen
  \bibfield  {author} {\bibinfo {author} {\bibfnamefont {J.}~\bibnamefont
  {Jordan}}, \bibinfo {author} {\bibfnamefont {R.}~\bibnamefont {Or\'us}},
  \bibinfo {author} {\bibfnamefont {G.}~\bibnamefont {Vidal}}, \bibinfo
  {author} {\bibfnamefont {F.}~\bibnamefont {Verstraete}},\ and\ \bibinfo
  {author} {\bibfnamefont {J.~I.}\ \bibnamefont {Cirac}},\ }\bibfield  {title}
  {\bibinfo {title} {{Classical Simulation of Infinite-Size Quantum Lattice
  Systems in Two Spatial Dimensions}},\ }\href
  {https://doi.org/10.1103/PhysRevLett.101.250602} {\bibfield  {journal}
  {\bibinfo  {journal} {Phys. Rev. Lett.}\ }\textbf {\bibinfo {volume} {101}},\
  \bibinfo {pages} {250602} (\bibinfo {year} {2008})}\BibitemShut {NoStop}%
\bibitem [{\citenamefont {Cirac}\ \emph {et~al.}(2021)\citenamefont {Cirac},
  \citenamefont {P\'erez-Garc\'{\i}a}, \citenamefont {Schuch},\ and\
  \citenamefont {Verstraete}}]{Cirac2021RMP}%
  \BibitemOpen
  \bibfield  {author} {\bibinfo {author} {\bibfnamefont {J.~I.}\ \bibnamefont
  {Cirac}}, \bibinfo {author} {\bibfnamefont {D.}~\bibnamefont
  {P\'erez-Garc\'{\i}a}}, \bibinfo {author} {\bibfnamefont {N.}~\bibnamefont
  {Schuch}},\ and\ \bibinfo {author} {\bibfnamefont {F.}~\bibnamefont
  {Verstraete}},\ }\bibfield  {title} {\bibinfo {title} {{Matrix Product States
  and Projected Entangled Pair States: Concepts, Symmetries, Theorems}},\
  }\href {https://doi.org/10.1103/RevModPhys.93.045003} {\bibfield  {journal}
  {\bibinfo  {journal} {Rev. Mod. Phys.}\ }\textbf {\bibinfo {volume} {93}},\
  \bibinfo {pages} {045003} (\bibinfo {year} {2021})}\BibitemShut {NoStop}%
\bibitem [{\citenamefont {Corboz}\ and\ \citenamefont
  {Vidal}(2009)}]{Corboz2009Fermionic}%
  \BibitemOpen
  \bibfield  {author} {\bibinfo {author} {\bibfnamefont {P.}~\bibnamefont
  {Corboz}}\ and\ \bibinfo {author} {\bibfnamefont {G.}~\bibnamefont {Vidal}},\
  }\bibfield  {title} {\bibinfo {title} {Fermionic multiscale entanglement
  renormalization ansatz},\ }\href {https://doi.org/10.1103/PhysRevB.80.165129}
  {\bibfield  {journal} {\bibinfo  {journal} {Phys. Rev. B}\ }\textbf {\bibinfo
  {volume} {80}},\ \bibinfo {pages} {165129} (\bibinfo {year}
  {2009})}\BibitemShut {NoStop}%
\bibitem [{\citenamefont {Corboz}\ \emph {et~al.}(2010)\citenamefont {Corboz},
  \citenamefont {Or\'us}, \citenamefont {Bauer},\ and\ \citenamefont
  {Vidal}}]{Corboz2010Simulation}%
  \BibitemOpen
  \bibfield  {author} {\bibinfo {author} {\bibfnamefont {P.}~\bibnamefont
  {Corboz}}, \bibinfo {author} {\bibfnamefont {R.}~\bibnamefont {Or\'us}},
  \bibinfo {author} {\bibfnamefont {B.}~\bibnamefont {Bauer}},\ and\ \bibinfo
  {author} {\bibfnamefont {G.}~\bibnamefont {Vidal}},\ }\bibfield  {title}
  {\bibinfo {title} {Simulation of strongly correlated fermions in two spatial
  dimensions with fermionic projected entangled-pair states},\ }\href
  {https://doi.org/10.1103/PhysRevB.81.165104} {\bibfield  {journal} {\bibinfo
  {journal} {Phys. Rev. B}\ }\textbf {\bibinfo {volume} {81}},\ \bibinfo
  {pages} {165104} (\bibinfo {year} {2010})}\BibitemShut {NoStop}%
\bibitem [{\citenamefont {Barthel}\ \emph {et~al.}(2009)\citenamefont
  {Barthel}, \citenamefont {Pineda},\ and\ \citenamefont
  {Eisert}}]{Barthel2009Contraction}%
  \BibitemOpen
  \bibfield  {author} {\bibinfo {author} {\bibfnamefont {T.}~\bibnamefont
  {Barthel}}, \bibinfo {author} {\bibfnamefont {C.}~\bibnamefont {Pineda}},\
  and\ \bibinfo {author} {\bibfnamefont {J.}~\bibnamefont {Eisert}},\
  }\bibfield  {title} {\bibinfo {title} {Contraction of fermionic operator
  circuits and the simulation of strongly correlated fermions},\ }\href
  {https://doi.org/10.1103/PhysRevA.80.042333} {\bibfield  {journal} {\bibinfo
  {journal} {Phys. Rev. A}\ }\textbf {\bibinfo {volume} {80}},\ \bibinfo
  {pages} {042333} (\bibinfo {year} {2009})}\BibitemShut {NoStop}%
\bibitem [{\citenamefont {Kraus}\ \emph {et~al.}(2010)\citenamefont {Kraus},
  \citenamefont {Schuch}, \citenamefont {Verstraete},\ and\ \citenamefont
  {Cirac}}]{Kraus2010Fermionic}%
  \BibitemOpen
  \bibfield  {author} {\bibinfo {author} {\bibfnamefont {C.~V.}\ \bibnamefont
  {Kraus}}, \bibinfo {author} {\bibfnamefont {N.}~\bibnamefont {Schuch}},
  \bibinfo {author} {\bibfnamefont {F.}~\bibnamefont {Verstraete}},\ and\
  \bibinfo {author} {\bibfnamefont {J.~I.}\ \bibnamefont {Cirac}},\ }\bibfield
  {title} {\bibinfo {title} {Fermionic projected entangled pair states},\
  }\href {https://doi.org/10.1103/PhysRevA.81.052338} {\bibfield  {journal}
  {\bibinfo  {journal} {Phys. Rev. A}\ }\textbf {\bibinfo {volume} {81}},\
  \bibinfo {pages} {052338} (\bibinfo {year} {2010})}\BibitemShut {NoStop}%
\bibitem [{\citenamefont {Corboz}\ \emph {et~al.}(2018)\citenamefont {Corboz},
  \citenamefont {Czarnik}, \citenamefont {Kapteijns},\ and\ \citenamefont
  {Tagliacozzo}}]{PhysRevX.8.031031}%
  \BibitemOpen
  \bibfield  {author} {\bibinfo {author} {\bibfnamefont {P.}~\bibnamefont
  {Corboz}}, \bibinfo {author} {\bibfnamefont {P.}~\bibnamefont {Czarnik}},
  \bibinfo {author} {\bibfnamefont {G.}~\bibnamefont {Kapteijns}},\ and\
  \bibinfo {author} {\bibfnamefont {L.}~\bibnamefont {Tagliacozzo}},\
  }\bibfield  {title} {\bibinfo {title} {{Finite Correlation Length Scaling
  with Infinite Projected Entangled-Pair States}},\ }\href
  {https://doi.org/10.1103/PhysRevX.8.031031} {\bibfield  {journal} {\bibinfo
  {journal} {Phys. Rev. X}\ }\textbf {\bibinfo {volume} {8}},\ \bibinfo {pages}
  {031031} (\bibinfo {year} {2018})}\BibitemShut {NoStop}%
\bibitem [{\citenamefont {Rams}\ \emph {et~al.}(2018)\citenamefont {Rams},
  \citenamefont {Czarnik},\ and\ \citenamefont {Cincio}}]{PhysRevX.8.041033}%
  \BibitemOpen
  \bibfield  {author} {\bibinfo {author} {\bibfnamefont {M.~M.}\ \bibnamefont
  {Rams}}, \bibinfo {author} {\bibfnamefont {P.}~\bibnamefont {Czarnik}},\ and\
  \bibinfo {author} {\bibfnamefont {L.}~\bibnamefont {Cincio}},\ }\bibfield
  {title} {\bibinfo {title} {{Precise Extrapolation of the Correlation Function
  Asymptotics in Uniform Tensor Network States with Application to the
  Bose-Hubbard and XXZ Models}},\ }\href
  {https://doi.org/10.1103/PhysRevX.8.041033} {\bibfield  {journal} {\bibinfo
  {journal} {Phys. Rev. X}\ }\textbf {\bibinfo {volume} {8}},\ \bibinfo {pages}
  {041033} (\bibinfo {year} {2018})}\BibitemShut {NoStop}%
\bibitem [{\citenamefont {Rader}\ and\ \citenamefont
  {L\"auchli}(2018)}]{PhysRevX.8.031030}%
  \BibitemOpen
  \bibfield  {author} {\bibinfo {author} {\bibfnamefont {M.}~\bibnamefont
  {Rader}}\ and\ \bibinfo {author} {\bibfnamefont {A.~M.}\ \bibnamefont
  {L\"auchli}},\ }\bibfield  {title} {\bibinfo {title} {{Finite Correlation
  Length Scaling in Lorentz-Invariant Gapless iPEPS Wave Functions}},\ }\href
  {https://doi.org/10.1103/PhysRevX.8.031030} {\bibfield  {journal} {\bibinfo
  {journal} {Phys. Rev. X}\ }\textbf {\bibinfo {volume} {8}},\ \bibinfo {pages}
  {031030} (\bibinfo {year} {2018})}\BibitemShut {NoStop}%
\bibitem [{SM()}]{SM}%
  \BibitemOpen
  \href@noop {} {\bibinfo {title} {{In Supplementary Sec.~I, we provide the
  details and comparisons between the SU and FFU. In Secs.~II and III, we show
  the process for extrapolating the SC order parameters obtained from simple
  update to the infinite-$D$ limit, for pristine and rare-earth element
  substituted nickelate R$_3$Ni$_2$O$_7$. In Sec.~IV, we provide results for
  larger iPEPS unit cells and comparisons among them.}}}\BibitemShut {Stop}%
\bibitem [{\citenamefont {Jiang}\ \emph {et~al.}(2008)\citenamefont {Jiang},
  \citenamefont {Weng},\ and\ \citenamefont {Xiang}}]{Xiang2008SU}%
  \BibitemOpen
  \bibfield  {author} {\bibinfo {author} {\bibfnamefont {H.~C.}\ \bibnamefont
  {Jiang}}, \bibinfo {author} {\bibfnamefont {Z.~Y.}\ \bibnamefont {Weng}},\
  and\ \bibinfo {author} {\bibfnamefont {T.}~\bibnamefont {Xiang}},\ }\bibfield
   {title} {\bibinfo {title} {{Accurate Determination of Tensor Network State
  of Quantum Lattice Models in Two Dimensions}},\ }\href
  {https://doi.org/10.1103/PhysRevLett.101.090603} {\bibfield  {journal}
  {\bibinfo  {journal} {Phys. Rev. Lett.}\ }\textbf {\bibinfo {volume} {101}},\
  \bibinfo {pages} {090603} (\bibinfo {year} {2008})}\BibitemShut {NoStop}%
\bibitem [{\citenamefont {Li}\ \emph {et~al.}(2012)\citenamefont {Li},
  \citenamefont {von Delft},\ and\ \citenamefont {Xiang}}]{Li2012SU}%
  \BibitemOpen
  \bibfield  {author} {\bibinfo {author} {\bibfnamefont {W.}~\bibnamefont
  {Li}}, \bibinfo {author} {\bibfnamefont {J.}~\bibnamefont {von Delft}},\ and\
  \bibinfo {author} {\bibfnamefont {T.}~\bibnamefont {Xiang}},\ }\bibfield
  {title} {\bibinfo {title} {Efficient simulation of infinite tree tensor
  network states on the {Bethe} lattice},\ }\href
  {https://doi.org/10.1103/PhysRevB.86.195137} {\bibfield  {journal} {\bibinfo
  {journal} {Phys. Rev. B}\ }\textbf {\bibinfo {volume} {86}},\ \bibinfo
  {pages} {195137} (\bibinfo {year} {2012})}\BibitemShut {NoStop}%
\bibitem [{\citenamefont {Phien}\ \emph {et~al.}(2015)\citenamefont {Phien},
  \citenamefont {Bengua}, \citenamefont {Tuan}, \citenamefont {Corboz},\ and\
  \citenamefont {Or\'us}}]{FFU2015}%
  \BibitemOpen
  \bibfield  {author} {\bibinfo {author} {\bibfnamefont {H.~N.}\ \bibnamefont
  {Phien}}, \bibinfo {author} {\bibfnamefont {J.~A.}\ \bibnamefont {Bengua}},
  \bibinfo {author} {\bibfnamefont {H.~D.}\ \bibnamefont {Tuan}}, \bibinfo
  {author} {\bibfnamefont {P.}~\bibnamefont {Corboz}},\ and\ \bibinfo {author}
  {\bibfnamefont {R.}~\bibnamefont {Or\'us}},\ }\bibfield  {title} {\bibinfo
  {title} {Infinite projected entangled pair states algorithm improved: Fast
  full update and gauge fixing},\ }\href
  {https://doi.org/10.1103/PhysRevB.92.035142} {\bibfield  {journal} {\bibinfo
  {journal} {Phys. Rev. B}\ }\textbf {\bibinfo {volume} {92}},\ \bibinfo
  {pages} {035142} (\bibinfo {year} {2015})}\BibitemShut {NoStop}%
\bibitem [{\citenamefont {Corboz}\ \emph {et~al.}(2014)\citenamefont {Corboz},
  \citenamefont {Rice},\ and\ \citenamefont {Troyer}}]{Corboz2014Competing}%
  \BibitemOpen
  \bibfield  {author} {\bibinfo {author} {\bibfnamefont {P.}~\bibnamefont
  {Corboz}}, \bibinfo {author} {\bibfnamefont {T.~M.}\ \bibnamefont {Rice}},\
  and\ \bibinfo {author} {\bibfnamefont {M.}~\bibnamefont {Troyer}},\
  }\bibfield  {title} {\bibinfo {title} {{Competing States in the {$t$-$J$}
  Model: {Uniform} $d$-Wave State versus Stripe State}},\ }\href
  {https://doi.org/10.1103/PhysRevLett.113.046402} {\bibfield  {journal}
  {\bibinfo  {journal} {Phys. Rev. Lett.}\ }\textbf {\bibinfo {volume} {113}},\
  \bibinfo {pages} {046402} (\bibinfo {year} {2014})}\BibitemShut {NoStop}%
\bibitem [{\citenamefont {Or\'us}\ and\ \citenamefont
  {Vidal}(2009)}]{Orus2009Simulation}%
  \BibitemOpen
  \bibfield  {author} {\bibinfo {author} {\bibfnamefont {R.}~\bibnamefont
  {Or\'us}}\ and\ \bibinfo {author} {\bibfnamefont {G.}~\bibnamefont {Vidal}},\
  }\bibfield  {title} {\bibinfo {title} {Simulation of two-dimensional quantum
  systems on an infinite lattice revisited: Corner transfer matrix for tensor
  contraction},\ }\href {https://doi.org/10.1103/PhysRevB.80.094403} {\bibfield
   {journal} {\bibinfo  {journal} {Phys. Rev. B}\ }\textbf {\bibinfo {volume}
  {80}},\ \bibinfo {pages} {094403} (\bibinfo {year} {2009})}\BibitemShut
  {NoStop}%
\bibitem [{\citenamefont {Hirthe}\ \emph {et~al.}(2023)\citenamefont {Hirthe},
  \citenamefont {Chalopin}, \citenamefont {Bourgund}, \citenamefont
  {Bojovi{\'c}}, \citenamefont {Bohrdt}, \citenamefont {Demler}, \citenamefont
  {Grusdt}, \citenamefont {Bloch},\ and\ \citenamefont
  {Hilker}}]{Hilker2023pairing}%
  \BibitemOpen
  \bibfield  {author} {\bibinfo {author} {\bibfnamefont {S.}~\bibnamefont
  {Hirthe}}, \bibinfo {author} {\bibfnamefont {T.}~\bibnamefont {Chalopin}},
  \bibinfo {author} {\bibfnamefont {D.}~\bibnamefont {Bourgund}}, \bibinfo
  {author} {\bibfnamefont {P.}~\bibnamefont {Bojovi{\'c}}}, \bibinfo {author}
  {\bibfnamefont {A.}~\bibnamefont {Bohrdt}}, \bibinfo {author} {\bibfnamefont
  {E.}~\bibnamefont {Demler}}, \bibinfo {author} {\bibfnamefont
  {F.}~\bibnamefont {Grusdt}}, \bibinfo {author} {\bibfnamefont
  {I.}~\bibnamefont {Bloch}},\ and\ \bibinfo {author} {\bibfnamefont {T.~A.}\
  \bibnamefont {Hilker}},\ }\bibfield  {title} {\bibinfo {title} {Magnetically
  mediated hole pairing in fermionic ladders of ultracold atoms},\ }\href
  {https://doi.org/10.1038/s41586-022-05437-y} {\bibfield  {journal} {\bibinfo
  {journal} {Nature}\ }\textbf {\bibinfo {volume} {613}},\ \bibinfo {pages}
  {463} (\bibinfo {year} {2023})}\BibitemShut {NoStop}%
\bibitem [{\citenamefont {Bohrdt}\ \emph {et~al.}(2022)\citenamefont {Bohrdt},
  \citenamefont {Homeier}, \citenamefont {Bloch}, \citenamefont {Demler},\ and\
  \citenamefont {Grusdt}}]{Grusdt2022mixD}%
  \BibitemOpen
  \bibfield  {author} {\bibinfo {author} {\bibfnamefont {A.}~\bibnamefont
  {Bohrdt}}, \bibinfo {author} {\bibfnamefont {L.}~\bibnamefont {Homeier}},
  \bibinfo {author} {\bibfnamefont {I.}~\bibnamefont {Bloch}}, \bibinfo
  {author} {\bibfnamefont {E.}~\bibnamefont {Demler}},\ and\ \bibinfo {author}
  {\bibfnamefont {F.}~\bibnamefont {Grusdt}},\ }\bibfield  {title} {\bibinfo
  {title} {Strong pairing in mixed-dimensional bilayer antiferromagnetic mott
  insulators},\ }\href {https://doi.org/10.1038/s41567-022-01561-8} {\bibfield
  {journal} {\bibinfo  {journal} {Nature Physics}\ }\textbf {\bibinfo {volume}
  {18}},\ \bibinfo {pages} {651} (\bibinfo {year} {2022})}\BibitemShut
  {NoStop}%
\bibitem [{\citenamefont {Li}\ \emph {et~al.}(2011)\citenamefont {Li},
  \citenamefont {Ran}, \citenamefont {Gong}, \citenamefont {Zhao},
  \citenamefont {Xi}, \citenamefont {Ye},\ and\ \citenamefont {Su}}]{Li2011a}%
  \BibitemOpen
  \bibfield  {author} {\bibinfo {author} {\bibfnamefont {W.}~\bibnamefont
  {Li}}, \bibinfo {author} {\bibfnamefont {S.-J.}\ \bibnamefont {Ran}},
  \bibinfo {author} {\bibfnamefont {S.-S.}\ \bibnamefont {Gong}}, \bibinfo
  {author} {\bibfnamefont {Y.}~\bibnamefont {Zhao}}, \bibinfo {author}
  {\bibfnamefont {B.}~\bibnamefont {Xi}}, \bibinfo {author} {\bibfnamefont
  {F.}~\bibnamefont {Ye}},\ and\ \bibinfo {author} {\bibfnamefont
  {G.}~\bibnamefont {Su}},\ }\bibfield  {title} {\bibinfo {title} {Linearized
  tensor renormalization group algorithm for the calculation of thermodynamic
  properties of quantum lattice models},\ }\href
  {https://doi.org/10.1103/PhysRevLett.106.127202} {\bibfield  {journal}
  {\bibinfo  {journal} {Phys. Rev. Lett.}\ }\textbf {\bibinfo {volume} {106}},\
  \bibinfo {pages} {127202} (\bibinfo {year} {2011})}\BibitemShut {NoStop}%
\bibitem [{\citenamefont {Dong}\ \emph {et~al.}(2017)\citenamefont {Dong},
  \citenamefont {Chen}, \citenamefont {Liu},\ and\ \citenamefont
  {Li}}]{Dong2017}%
  \BibitemOpen
  \bibfield  {author} {\bibinfo {author} {\bibfnamefont {Y.-L.}\ \bibnamefont
  {Dong}}, \bibinfo {author} {\bibfnamefont {L.}~\bibnamefont {Chen}}, \bibinfo
  {author} {\bibfnamefont {Y.-J.}\ \bibnamefont {Liu}},\ and\ \bibinfo {author}
  {\bibfnamefont {W.}~\bibnamefont {Li}},\ }\bibfield  {title} {\bibinfo
  {title} {Bilayer linearized tensor renormalization group approach for thermal
  tensor networks},\ }\href {https://doi.org/10.1103/PhysRevB.95.144428}
  {\bibfield  {journal} {\bibinfo  {journal} {Phys. Rev. B}\ }\textbf {\bibinfo
  {volume} {95}},\ \bibinfo {pages} {144428} (\bibinfo {year}
  {2017})}\BibitemShut {NoStop}%
\bibitem [{\citenamefont {Li}\ \emph {et~al.}(2023)\citenamefont {Li},
  \citenamefont {Gao}, \citenamefont {He}, \citenamefont {Qi}, \citenamefont
  {Chen},\ and\ \citenamefont {Li}}]{tanTRG2023}%
  \BibitemOpen
  \bibfield  {author} {\bibinfo {author} {\bibfnamefont {Q.}~\bibnamefont
  {Li}}, \bibinfo {author} {\bibfnamefont {Y.}~\bibnamefont {Gao}}, \bibinfo
  {author} {\bibfnamefont {Y.-Y.}\ \bibnamefont {He}}, \bibinfo {author}
  {\bibfnamefont {Y.}~\bibnamefont {Qi}}, \bibinfo {author} {\bibfnamefont
  {B.-B.}\ \bibnamefont {Chen}},\ and\ \bibinfo {author} {\bibfnamefont
  {W.}~\bibnamefont {Li}},\ }\bibfield  {title} {\bibinfo {title} {Tangent
  space approach for thermal tensor network simulations of the {2D} {Hubbard}
  model},\ }\href {https://doi.org/10.1103/PhysRevLett.130.226502} {\bibfield
  {journal} {\bibinfo  {journal} {Phys. Rev. Lett.}\ }\textbf {\bibinfo
  {volume} {130}},\ \bibinfo {pages} {226502} (\bibinfo {year}
  {2023})}\BibitemShut {NoStop}%
\bibitem [{\citenamefont {White}(2009)}]{White2009METTS}%
  \BibitemOpen
  \bibfield  {author} {\bibinfo {author} {\bibfnamefont {S.~R.}\ \bibnamefont
  {White}},\ }\bibfield  {title} {\bibinfo {title} {Minimally entangled typical
  quantum states at finite temperature},\ }\href
  {https://doi.org/10.1103/PhysRevLett.102.190601} {\bibfield  {journal}
  {\bibinfo  {journal} {Phys. Rev. Lett.}\ }\textbf {\bibinfo {volume} {102}},\
  \bibinfo {pages} {190601} (\bibinfo {year} {2009})}\BibitemShut {NoStop}%
\bibitem [{\citenamefont {Stoudenmire}\ and\ \citenamefont
  {White}(2010)}]{Stoudenmire2010}%
  \BibitemOpen
  \bibfield  {author} {\bibinfo {author} {\bibfnamefont {E.~M.}\ \bibnamefont
  {Stoudenmire}}\ and\ \bibinfo {author} {\bibfnamefont {S.~R.}\ \bibnamefont
  {White}},\ }\bibfield  {title} {\bibinfo {title} {Minimally entangled typical
  thermal state algorithms},\ }\href
  {https://doi.org/10.1088/1367-2630/12/5/055026} {\bibfield  {journal}
  {\bibinfo  {journal} {New J. Phys.}\ }\textbf {\bibinfo {volume} {12}},\
  \bibinfo {pages} {055026} (\bibinfo {year} {2010})}\BibitemShut {NoStop}%
\bibitem [{\citenamefont {Czarnik}\ and\ \citenamefont
  {Dziarmaga}(2014)}]{Czarnik2014PEPO}%
  \BibitemOpen
  \bibfield  {author} {\bibinfo {author} {\bibfnamefont {P.}~\bibnamefont
  {Czarnik}}\ and\ \bibinfo {author} {\bibfnamefont {J.}~\bibnamefont
  {Dziarmaga}},\ }\bibfield  {title} {\bibinfo {title} {Fermionic projected
  entangled pair states at finite temperature},\ }\href
  {https://doi.org/10.1103/PhysRevB.90.035144} {\bibfield  {journal} {\bibinfo
  {journal} {Phys. Rev. B}\ }\textbf {\bibinfo {volume} {90}},\ \bibinfo
  {pages} {035144} (\bibinfo {year} {2014})}\BibitemShut {NoStop}%
\bibitem [{\citenamefont {Czarnik}\ \emph {et~al.}(2016)\citenamefont
  {Czarnik}, \citenamefont {Rams},\ and\ \citenamefont
  {Dziarmaga}}]{Czarnik2016PRB}%
  \BibitemOpen
  \bibfield  {author} {\bibinfo {author} {\bibfnamefont {P.}~\bibnamefont
  {Czarnik}}, \bibinfo {author} {\bibfnamefont {M.~M.}\ \bibnamefont {Rams}},\
  and\ \bibinfo {author} {\bibfnamefont {J.}~\bibnamefont {Dziarmaga}},\
  }\bibfield  {title} {\bibinfo {title} {Variational tensor network
  renormalization in imaginary time: Benchmark results in the {Hubbard} model
  at finite temperature},\ }\href {https://doi.org/10.1103/PhysRevB.94.235142}
  {\bibfield  {journal} {\bibinfo  {journal} {Phys. Rev. B}\ }\textbf {\bibinfo
  {volume} {94}},\ \bibinfo {pages} {235142} (\bibinfo {year}
  {2016})}\BibitemShut {NoStop}%
\end{thebibliography}%


\newpage
\clearpage
\onecolumngrid
\mbox{}
\begin{center}
	
	{\large Supplementary Materials for} $\,$ \\
	\bigskip
	\textbf{\large{Orbital-selective Superconductivity in the Pressurized Bilayer Nickelate La$_3$Ni$_2$O$_7$: 
			\\ An Infinite Projected Entangled-Pair State Study}} \\
	Chen \textit{et al}.
\end{center}

\date{\today}

\setcounter{section}{0}
\setcounter{figure}{0}
\setcounter{equation}{0}
\renewcommand{\theequation}{S\arabic{equation}}
\renewcommand{\thefigure}{S\arabic{figure}}
\setcounter{secnumdepth}{3}

\section{Simple vs. full update in the iPEPS calculations}
\begin{figure*}[!htbp]
	\includegraphics[width=0.6\linewidth]{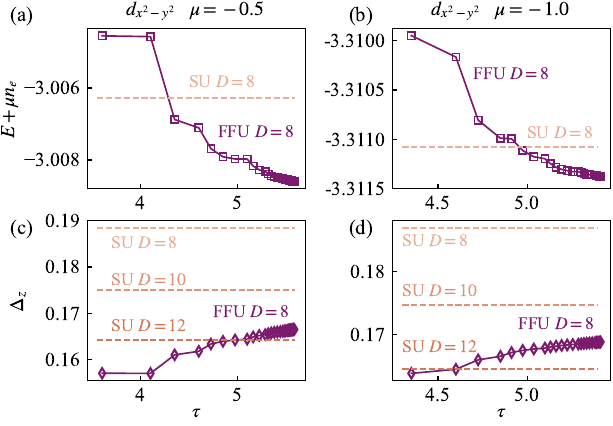}
	\caption{The FFU convergence process of (a, b) energy $E+\mu n_e$ and (c, d) 
		SC order parameter $\Delta_z$ with imaginary time $\tau$ for $\mu=-0.5$ (left) 
		and $\mu=-1.0$ (right). Red dashed lines represent results of SU with different 
		$D$, and open squares or diamonds for FFU with fixed $D=8$. The model parameters 
		are set as $t_\parallel=1$, $J_\parallel=1/3$, $t_\perp=0$, and $J_\perp=2/3$
		for \XO orbital. 
	}
	\label{Appendix_Convergence_x2}
\end{figure*}

We show in Fig.~\ref{Appendix_Convergence_x2} two representative convergence 
processes of our fast full update (FFU), as compared to the results of simply update 
(SU). FFU is more accurate than SU but with higher computation complexity, so its 
bond dimension $D$ is limited to 8 and 10 in the present study. In our calculations, a chemical 
potential  term $\mu n_e$  is added to Hamiltonian~(1) to control the 
electron density $n_e$. Chemical potential $\mu=-0.5$ and $-1.0$ correspond to 
the two adjacent points just beside $n_e=0.6$ (green dashed line) for the $d_{x^2-y^2}$ 
orbital in Fig.~2(a). In the update process, the imaginary time evolution 
operator $\exp[-(H+\mu n_e)\Delta\tau]$ with gradually decreasing $\Delta\tau$ 
(e.g., from 0.2 to 0.0005) acts on a randomly initialized state (for SU) or a saved 
stated (for FFU) obtained from, e.g., previous SU calculations. As shown in panels 
(a) and (b), the final energy $E+\mu n_e$ is converged and lower than that of SU 
for both $\mu=-0.5$ and $\mu=-1.0$ with the same bond dimension $D=8$. As 
shown in panels (c) and (d), the SC order parameter $\Delta_z$ of FFU with $D=8$ 
is even close to that of SU with larger $D$, showing the superior performance of 
FFU and the agreements between two update schemes. 

\section{Data Extrapolations with various $J_\perp$ and $t_\perp$}
\begin{figure*}[!htbp]
	\includegraphics[width=0.8\linewidth]{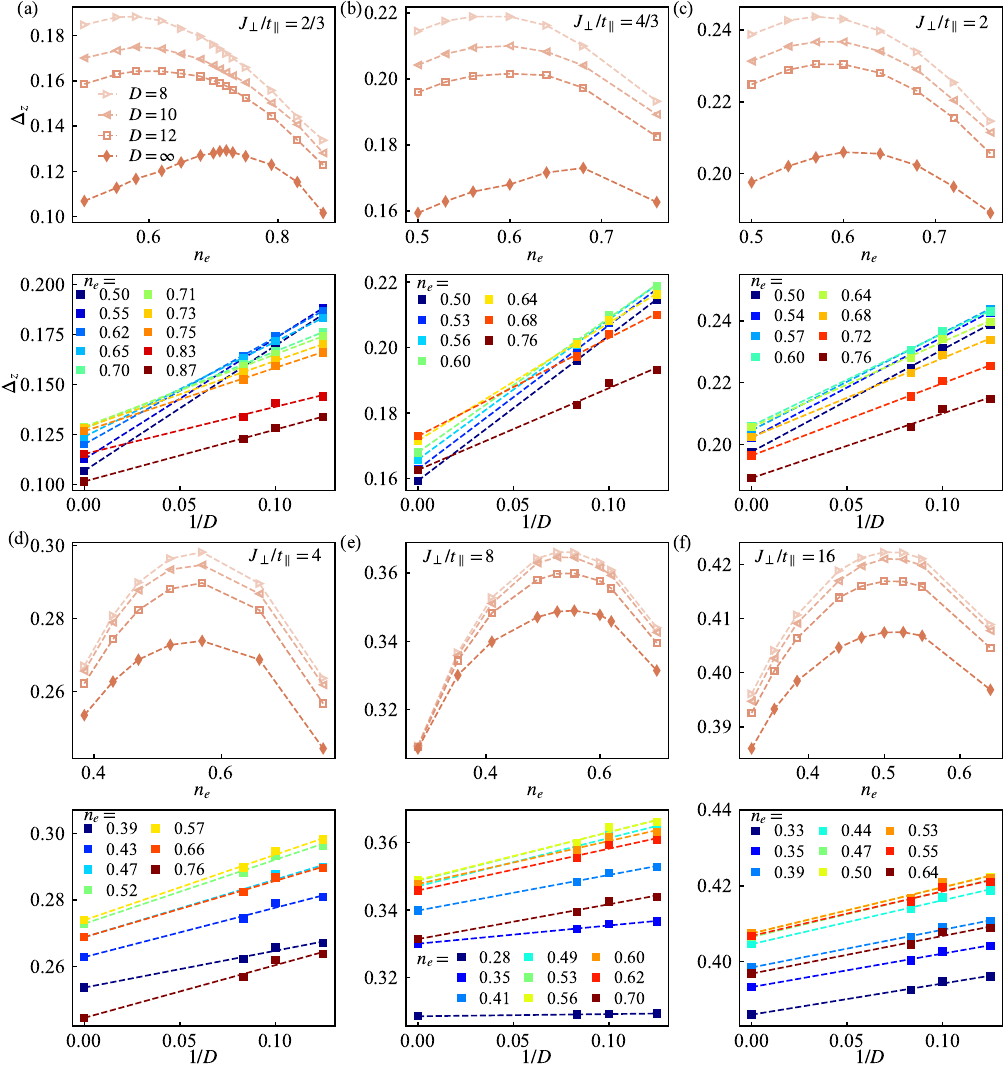}
	\caption{The SC order parameter $\Delta_z$ vs. electron 
		densities $n_e$ of the $d_{x^2-y^2}$ orbitals with 
		(a) $J_\perp/t_\parallel=2/3$, (b) $J_\perp/t_\parallel=4/3$, 
		(c) $J_\perp/t_\parallel=2$, (d) $J_\perp/t_\parallel=4$, 
		(e) $J_\perp/t_\parallel=8$, and (f) $J_\perp/t_\parallel=16$, 
		respectively. The lower panels show the linear extrapolation of 
		$\Delta_z$ with inverse bond dimension $1/D$, and the different 
		colors represent different densities $n_e$. Other model parameters 
		are fixed as $t_\parallel=1$, $J_\parallel=1/3$, and $t_\perp=0$.
	}
	\label{Appendix_Jperp}
\end{figure*}

We show in Fig.~\ref{Appendix_Jperp} the process to extrapolate the interlayer 
SC order parameter $\Delta_z$ to the infinite $D$ limit, which has been shown
in Fig.~2(a) and Fig.~3(d) of the main article. The SC order 
parameters $\Delta_z$ for $J_\perp/t_\parallel=2/3$, $4/3$, 2, 4, 8, and 16 
with finite bond dimension $D=8$, 10, 12 are plotted in panels (a-f), which are 
fitted with a linear function of $1/D$ and extrapolated to the infinite $D$ limit 
in the panels just below. In Fig.~\ref{Appendix_Jperp}, we find $\Delta_z$ gets 
enhanced by increasing $J_\perp$ and the optimal electron density $n_e$ 
shifts towards $n_e=0.5$, indicating a BCS-BEC crossover in this system.

\begin{figure*}[!htbp]
	\includegraphics[width=0.95\linewidth]{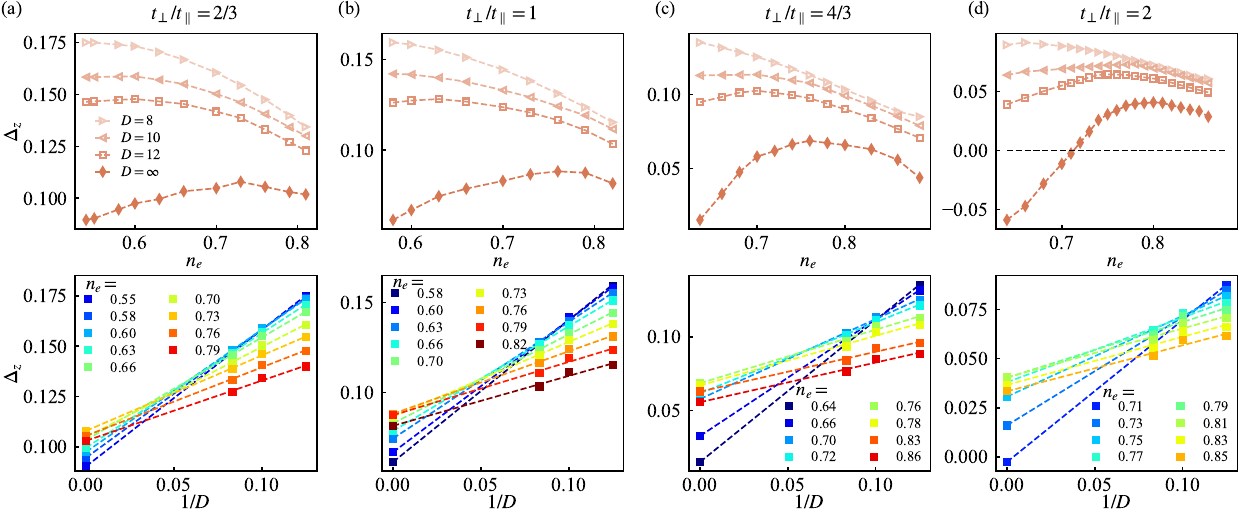}
	\caption{The SC order parameters $\Delta_z$ vs. electron densities $n_e$ 
		for the $d_{x^2-y^2}$ orbital with (a) $t_\perp/t_\parallel=2/3$, 
		(b) $t_\perp/t_\parallel=1$, (c) $t_\perp/t_\parallel=4/3$, and 
		(d) $t_\perp/t_\parallel=2$, respectively. The lower panels show
		the linear extrapolation of $\Delta_z$ with inverse bond dimension $1/D$, 
		and the different colors represent different densities $n_e$. Other model 
		parameters are fixed as $t_\parallel=1$, $J_\parallel=1/3$, and $J_\perp=2/3$.
	}
	\label{Appendix_tperp}
\end{figure*}

We show in Fig.~\ref{Appendix_tperp} the process to get extrapolated 
$\Delta_z$ at infinite $D$ limit in Fig.~3(a) of the main article. 
The SC order parameters $\Delta_z$ for $t_\perp/t_\parallel=2/3$, 4/3, 
2, 4, 8, and 16 with finite bond dimension $D=8$, 10, 12 are plotted in 
panels (a-d), and are extrapolated linearly with $1/D$ to the infinite $D$ 
limit in the panels just below. We can see that the SC order $\Delta_z$ 
gets suppressed by increasing $t_\perp$ and the optimal density $n_e$ 
shifts towards half filling, i.e., the low-doping regime.

\section{Data Extrapolations for RE element substitution}

\begin{figure*}[!htbp]
	\includegraphics[width=0.75\linewidth]{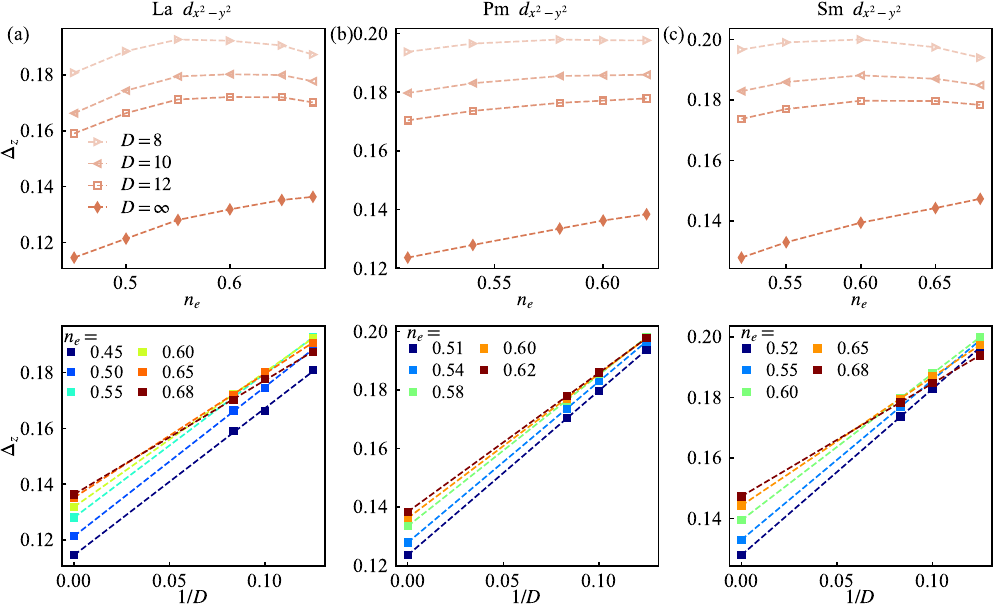}
	\caption{The SC order parameters $\Delta_z$ for various densities $n_e$ for 
		$d_{x^2-y^2}$ electrons with element substitution (a) La, (b) Pm, and (c) Sm. 
		The panels in the lower row show the extrapolation of $\Delta_z$ to $1/D=0$. 
		The model parameters follow those in Fig.~4(a) of the main
		article.}
	\label{Appendix_substitution_x}
\end{figure*}

We show in Fig.~\ref{Appendix_substitution_x} the process to extrapolate 
$\Delta_z$ in the $d_{x^2-y^2}$ orbital to the infinite $D$ limit, which has
been shown in Fig.~4(b) of the main article. The SC order parameters 
$\Delta_z$ for substitution of element La, Pm, and Sm with finite bond 
dimension $D=8$, 10, 12 are plotted in the upper row of panels (a-d), 
which are extrapolated linearly with $1/D$ to infinite $D$ limit in the lower
row of those panels. From the results, we find that the order parameter
$\Delta_z$ gets increased with substitution of La by Pm or Sm.

\section{Results for different iPEPS unit cells} 
\begin{figure*}[!htbp]
	\includegraphics[width=1.0\linewidth]{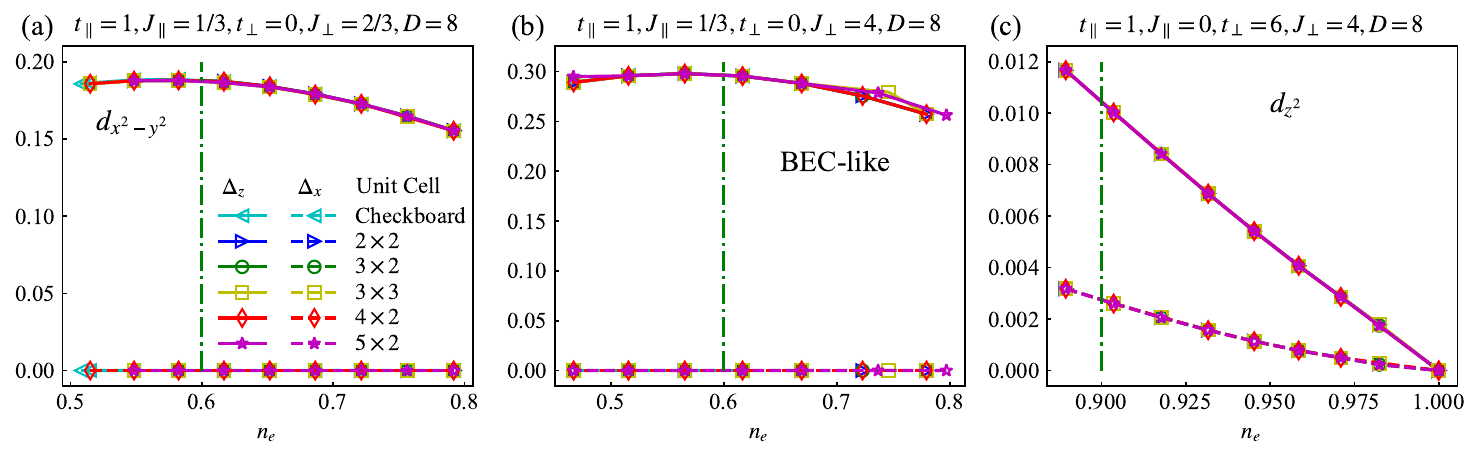}
	\caption{The SC order parameters $\Delta_z$ for the interlayer pairing and $\Delta_x$ for the intralayer pairing calculated with varying unit cell sizes for three different parameter representing (a) $d_{x^2-y^2}$ orbital, (b) $d_{x^2-y^2}$ orbital with larger $J_\perp=4$ (BEC-like case), and (c) $d_{z^2}$ orbital. The bond dimension $D=8$ in all calculations. The legend of (b) and (c) is the same as that shown in (a). The green vertical lines mark different electron densities in the $d_{x^2-y^2}$ and $d_{z^2}$ orbitals, where $n_{x^2-y^2} \simeq 0.6$ and $n_{z^2}\simeq 0.9$ in~\LNO. The model parameters are $t_\parallel=1$, $J_\parallel=1/3$, $t_\perp=0$, $J_\perp=2/3$ for the $d_{x^2-y^2}$ orbital in (a), $t_\parallel=1$, $J_\parallel=1/3$, $t_\perp=0$,  $J_\perp=4$ in (b), and $t_\parallel=1/6$, $J_\parallel=0$, $t_\perp=1$, $J_\perp=2/3$ for the \ZO orbital in (c). }
	\label{Appendix_pair_diffUC}
\end{figure*}

 In Fig.~\ref{Appendix_pair_diffUC} we show results obtained with different unit cells of size 
	$N_x \times N_y = $ $2\times 2$, $3\times 2$, $3\times 3$, $4\times 2$, $5\times 2$.
	As shown in these figures, the SC order parameters $\Delta_z$ of the interlayer 
	pairing and $\Delta_x$ for the intralayer pairing do not change with different unit cells. 
	Our study reveals that SC order is notably resilient within the \XO 
	orbital and the modified \XO orbital with an increased exchange interaction (e.g.,
	$J_\perp  =4$). Conversely, the $d_{z^2}$ orbital exhibits only a faint trace of SC order, which 
	remains unaltered regardless of the chosen unit cell configurations. Moreover, the charge 
	distribution is found to be homogeneous throughout the system, and the magnitude of 
	magnetic moments is vanishingly small, thus indicating an absence of competing 
	charge or spin ordering in the ground state for the parameters under consideration.

\end{document}